\documentclass[reprint, NumberedRefs]{JASA}

 \usepackage{amsmath,amssymb,amsfonts}
\usepackage{array}
\usepackage[caption=false,font=footnotesize]{subfig}
\usepackage{url}
\usepackage{stmaryrd}
\usepackage{tcolorbox}   
\usepackage{graphicx,graphics,color,psfrag}

\newcommand{\bc}[1]{\mbox{\boldmath $\mathcal{#1}$}}
\newcommand{\mb}[1]{\mathbf{#1}}
\newcommand{\F}{\mathrm{F}}
\newcommand{\T}{\mathrm{T}}

\begin{document}

\title[JASA/Sample JASA Article]{Tensor-based Basis Function Learning for Three-dimensional  Sound Speed Fields}
\author{Lei Cheng}
\author{Xingyu Ji}
\affiliation{College of Information Science and Electronic Engineering, Zhejiang University, Hangzhou, 310027, China}

\author{Hangfang Zhao}

\affiliation{College of Information Science and Electronic Engineering, Zhejiang University, Hangzhou, 310027, China}

\affiliation{Zhejiang Provincial Key Laboratory of Ocean Observation-Imaging Testbed, Ocean College, Zhejiang University, Zhoushan, 316000, China}

\affiliation{The Engineering Research Center of Oceanic Sensing Technology and Equipment, Ministry of Education}

\author{Jianlong Li}

\affiliation{College of Information Science and Electronic Engineering, Zhejiang University, Hangzhou, 310027, China}

\affiliation{Zhejiang Provincial Key Laboratory of Ocean Observation-Imaging Testbed, Ocean College, Zhejiang University, Zhoushan, 316000, China}

\affiliation{The Engineering Research Center of Oceanic Sensing Technology and Equipment, Ministry of Education}

\author{Wen Xu}
\affiliation{College of Information Science and Electronic Engineering, Zhejiang University, Hangzhou, 310027, China}

\affiliation{Zhejiang Provincial Key Laboratory of Ocean Observation-Imaging Testbed, Ocean College, Zhejiang University, Zhoushan, 316000, China}

\affiliation{The Engineering Research Center of Oceanic Sensing Technology and Equipment, Ministry of Education}

\preprint{Author, JASA}	

\date{\today}

\begin{abstract}
Basis function learning is the stepping stone towards effective three-dimensional (3D) sound speed field (SSF) inversion for various acoustic signal processing tasks, including ocean acoustic tomography, underwater  target localization/tracking, and underwater communications. Classical basis functions include the empirical orthogonal functions (EOFs), Fourier basis functions, and their combinations. The unsupervised machine learning method, e.g., the K-SVD algorithm, has recently tapped into the basis function design, showing better representation performance than the EOFs. {\color{black} However, existing methods do not consider basis function learning  approaches that treat 3D SSF data as a third-order tensor}, and thus cannot fully utilize the 3D interactions/correlations therein. To circumvent such a drawback, basis function learning is linked to tensor decomposition in this paper, which is the primary drive for recent multi-dimensional data mining.  In particular, a  tensor-based basis function learning framework is proposed, which can include the classical basis functions (using EOFs and/or Fourier basis functions) as its special cases. This provides a unified tensor perspective for understanding and representing 3D SSFs. Numerical results using the South China Sea 3D SSF data have demonstrated the excellent performance of the tensor-based basis functions.
\end{abstract}


\maketitle


\section{\label{sec:1} Introduction}

Basis function learning for sound speed fields (SSFs) has played a vital role in a wide range of  acoustic signal processing tasks, such as ocean acoustic tomography \cite{munk2009ocean, zhu2020}, underwater target localization/tracking \cite{li2011time, michalopoulou2021matched}, and underwater communications\cite{qu2014two}.  {\color{black} Effective basis functions}  can significantly reduce the number of unknown parameters to be estimated, thereby making the originally under-determined SSF inversion problem much more manageable. The underlying rationale is that sound speeds are correlated across spatial and temporal domains\cite{munk2009ocean, zhu2020, bianco2016compressive, huang2014method}, making SSFs viable to be accurately represented by a set of basis functions. {\color{black} These basis functions are expected to} have high expressive power such that only a few of them are capable of accurate SSF representation. 

Recently, this goal {\color{black} was noticed to coincide} with the aim of unsupervised representation learning\cite{bianco2018machine}, and thus has triggered the surging development of machine learning \cite{theodoridis2020machine} for ocean {\color{black} acoustics} \cite{bianco2019machine, niu2021mode, ozanich2020feedforward}. Specifically, the classical empirical orthogonal functions (EOFs)\cite{leblanc1980underwater} can be interpreted as the basis vectors derived from principal component analysis (PCA)\cite{wold1987principal}, which suggests that the nonlinear variants of PCA, e.g., kernel PCA \cite{scholkopf1997kernel}, can potentially give nonlinear SSF representations. Furthermore, popular dictionary learning (DL) methods\cite{tovsic2011dictionary}, e.g., K-SVD\cite{aharon2006k}, which have shown remarkable performance  in the image and video de-noising, were introduced to learn a {\color{black} reduced-order} representation of SSFs \cite{bianco2017dictionary}, showing improved generalization performance in SSF reconstruction. The success behind EOFs and K-SVD-based approach lies in that the basis functions are directly learnt from the training SSF data\cite{elad2010sparse}, {\color{black} while} not generated {\it ad hoc} from a standard set of functions such as Fourier basis functions \cite{bracewell1986fourier} or wavelets \cite{antonini1992image}, {\color{black} which} exemplifies the effectiveness of  data-driven approach in representation learning. 

 \begin{figure*}[t]
\includegraphics[width= 6in]{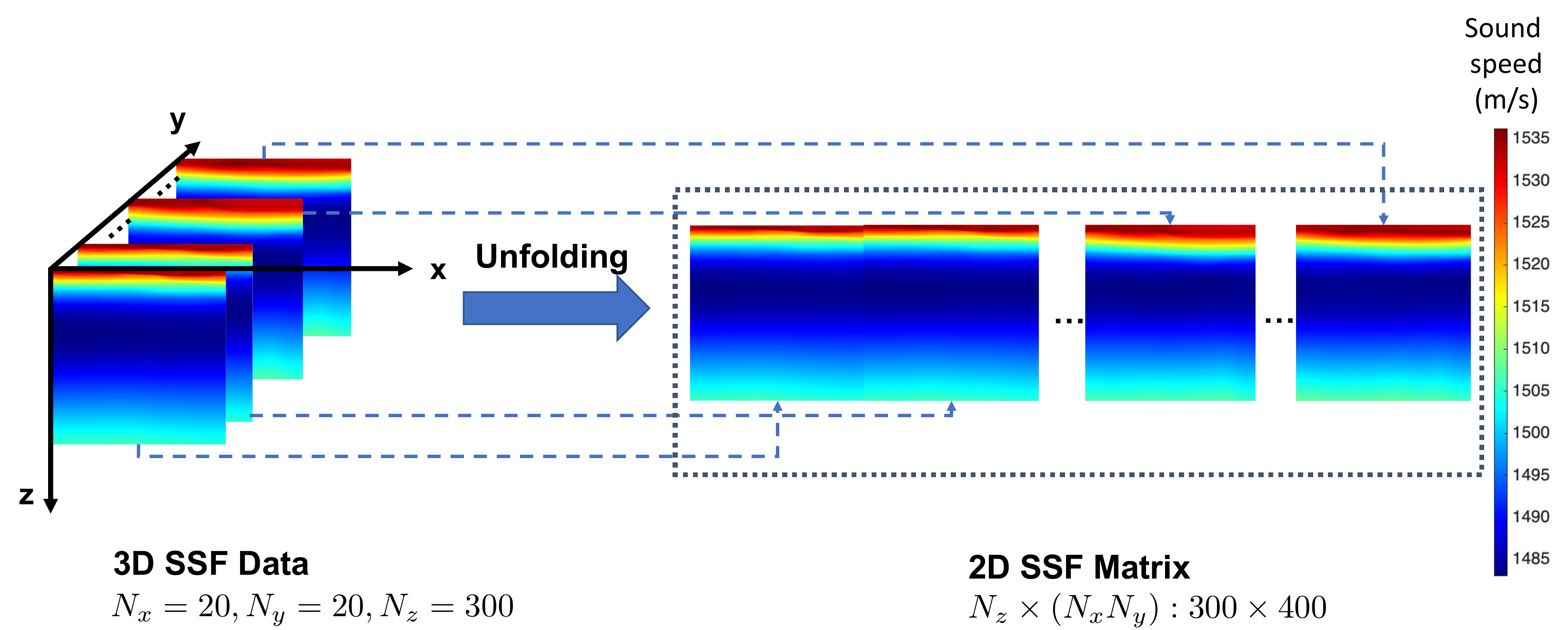}
\caption{\label{fig:FIG1}{Illustration of three-dimensional (3D) sound speed field (SSF) data and its matrix unfolding operation.}}
\raggedright
\hrule
\end{figure*}

 For 3D SSF data, as illustrated in Fig.~\ref{fig:FIG1}, {\color{black} sound speeds} are correlated across 3D coordinates, {\color{black} since the processes driving the ocean sound speed profiles are inherently continuous in space and time.}  On the other hand, both EOFs and K-SVD algorithm are designed for two-dimensional (2D) data, e.g., an SSF matrix, and thus did not take the multi-dimensional correlations among sound speeds into account. Therefore, to use EOFs or K-SVD method, we should firstly unfold the 3D SSF data into an SSF matrix (see Fig.~\ref{fig:FIG1}). However, the matrix unfolding operation (also called matricization)  {\color{black}breaks} the original 3D structure of data and thus induce information loss\cite{sidiropoulos2017tensor, panagakis2021tensor}. This drawback has been theoretically proved in the multi-dimensional harmonic retrieval task \cite{roemer2014analytical}, and reported in various signal processing applications, including directional-of-arrival (DOA) estimation \cite{cheng2015subspace}, blind source separation \cite{cheng2020learning} and image completion\cite{zhao2015bayesian}. This difficulty leads to an immediate question: {\it how to avoid the information loss caused by matricization in a principled manner? }

This question invites the framework of {\it tensor decomposition} and the associated {\it multi-linear algebra}  \cite{kolda2009tensor} , which are much richer than their matrix-based counterparts. Over the past two decades, tensor decomposition has become the primary drive for understanding/representing  multi-dimensional data, and has achieved great success in various machine learning and signal processing applications\cite{sidiropoulos2017tensor, panagakis2021tensor}. 
In this paper, we not only show the state-of-the-art (SOTA) performance of tensor-based basis function learning for 3D SSFs via extensive numerical results,  but also theoretically prove that the classical basis functions (using EOFs and/or Fourier basis functions) \cite{leblanc1980underwater, cornuelle1989ocean, morawitz1996three} are interestingly the special case of the proposed tensor-based learning framework. The latter insight justifies the effectiveness of the tensor-based approach from another angle, and further paves the way for future investigation of better basis functions through the lens of a unified tensor perspective.

The remainder of this paper is organized as follows. In Section~\ref{sec:2}, we briefly review  classical basis functions as well as the recent ones for SSF representation. In Section~\ref{sec:3}, 
representation learning using tensor decomposition is introduced. Then, we propose a tensor-based basis function learning framework for 3D SSF, and further reveal the connections between the classical basis functions and the tensor-based counterparts in  Section~\ref{sec:4}. Tensor-based basis function learning algorithms for one 3D SSF and multiple 3D SSFs are introduced in  Section~\ref{sec:4}. Extensive numerical results are reported in Section~\ref{sec:5}. Finally, conclusions and future research directions are discussed in Section~\ref{sec:6}.  

{\it Notations:} Lower- and upper-case bold letters are used to denote vectors and matrices, respectively. Higher-order tensors are denoted by upper-case bold calligraphic letters. For a tensor $\bc{X}$, $\mb{X}_{(p)}$ stands for its mode-$p$ unfolding matrix. $\bc X \times_p \mb B$ denotes the $p$-mode product between tensor $\bc X$ and matrix $\mb B$. The Kronecker product is denoted by $\otimes$. The superscripts $^{\T}$ and  $^{\text{H}}$ stand for transposition and Hermitian respectively. $^\dagger$ denotes the Moore-Penrose pseudo inverse. $\mathrm{diag}(\mb{x})$ denotes the diagonal matrix with $\mb{x}$ on its main diagonal. The identity matrix of order $N$ is denoted by $\mb{I}_N$. $\|\cdot\|_{\F}$ stands for the Frobenius norm. $\mathbb{R}$ and $\mathbb{C}$ are the field of real numbers and complex numbers, respectively.

\section{\label{sec:2} Basis Functions for Sound Speed Fields}

\subsection{\label{Sec II-A} 2D SSF: PCA and EOFs}

Consider a 2D SSF matrix $\mb Y = [\mb y_1, \cdots, \mb y_J] \in \mathbb R^{I \times J}$ {\color{black} containing}  $J$ 1D temporal/spatial sound speed profiles (SSPs),  where $I$ {\color{black} usually denotes} the number of discrete points in depth. 

In order to extract the basis functions that capture the most variances of data, PCA\cite{wold1987principal} is performed on the SSF matrix $\mb Y$. Particularly, $\mb Y$ is first centered by subtracting a mean matrix $\mb M = [\mb m, \cdots, \mb m] \in \mathbb R^{I \times J}$ with $\mb m = \frac{1}{J} \sum_{j=1}^J \mb y_{j} \in \mathbb R^{I\times 1}$, giving a zero-mean SSF matrix $\mb X = [\mb x_1, \cdots, \mb x_J] = \mb Y - \mb M \in \mathbb R^{I \times J}$, in which each element is known as SSF perturbation \cite{munk2009ocean, zhu2020}.  Then, the eigenvalue decomposition (EVD) of the correlation matrix $\mb X \mb X^\T$ finds the EOFs \cite{leblanc1980underwater} as follows:
\begin{align}
     \mb X \mb X^\T = \mb E \mb \Lambda \mb E^\T,
\end{align}
where $\mb E = [\mb e_1, \cdots, \mb e_I]$ {\color{black} contains} the EOFs $\{\mb e_i\}_i$ (eigenvectors) and $\mb \Lambda = \text{diag}([\lambda_1, \cdots, \lambda_I])$  {\color{black} contains the eigenvalue} $\lambda_i$ associated with the $i$-th EOF $\mb e_i$, for $i=1,\cdots, I$. Without loss of generality, it is assumed that $\lambda_1 \geq \cdots \geq \lambda_I$.

Typically, we only retain $K (K \leq I)$ leading-order EOFs to represent the SSF matrix for dimensionality reduction. Given the EOF matrix $\mb E_K = [\mb e_1, \cdots, \mb e_K] \in \mathbb R^{I \times K}$ , the zero-mean SSF matrix can be approximately represented by  \cite{leblanc1980underwater}
\begin{align}
    \mb X \approx \mb E_K \mb W,
\end{align}
where $\mb W \in \mathbb R^{K \times J}$ is the representation coefficient matrix. Since the EOF matrix $\mb E_K$  is orthonormal (i.e., $\mb E_K^{\text{T}}\mb E_K  = \mb I_K$), the least-squares (LS) estimates of the coefficient matrix  can be efficiently computed  by  \cite{leblanc1980underwater}
\begin{align}
 \hat{\mb W}  = \mb E_K^{\text{T}} \mb X.
\end{align}

For an unseen zero-mean SSF sample $\mb X^* \in \mathbb R^{I \times J^\prime}$, given EOF matrix $\mb E_K$,  {\color{black} the} coefficient matrix $\mb W^* \in \mathbb R^{K \times J^\prime}$  used for SSF representation is computed by  \cite{leblanc1980underwater}
\begin{align}
\mb W^* =  \mb E_K^{\text{T}} \mb X^*.
\end{align}

\subsection{2D SSF: K-SVD and Over-complete Dictionary}
To seek a more  {\color{black} effective reduced-order} representation of SSF,  dictionary learning methods\cite{tovsic2011dictionary, aharon2006k}, which were originally designed for image/video de-noising,  has recently tapped into ocean  {\color{black}signal processing} \cite{bianco2017dictionary}. The key idea is to jointly optimize an over-complete dictionary matrix $\mb Q \in \mathbb R^{I \times Z} ( Z \geq I)$ and the associated sparse coefficient matrix $\mb V$ such that the reconstruction error is minimized \cite{aharon2006k, bianco2017dictionary}:
\begin{align}
&\min_{\mb Q \in \mathbb R^{I \times Z}} \left \{ \min_{\mb V \in \mathbb R^{Z \times J}} ||\mb X - \mb Q \mb V ||_\F^2 \right\},\nonumber \\
& \text{s.t.}~~  ||\mb V_{:,j} ||_0 \leq T, ~~ j = 1,\cdots J,
\label{opt:dl}
\end{align}
where $T$ is a pre-defined upper bound value for the number of non-zero elements in each column $\mb V_{:,j}, \forall j$.

To solve  {\color{black} the} problem  {\color{black} in Eq.~\eqref{opt:dl}} in a computationally efficient manner, K-SVD \cite{aharon2006k} was proposed to alternatively  update the dictionary matrix $\mb Q$ (called dictionary update step) and the coefficient matrix $\mb V$ (called sparse coding step). More concretely, in the $t$-th iteration, given dictionary matrix $\mb Q^{t-1}$ and denote the $j$-th column of matrix $\mb V$ as $\mb v_j$, sparse coding step consists of $J$ subproblems  \cite{aharon2006k}:
\begin{align}
&\min_{\mb v_j} || \mb x_j - \mb Q^{t-1} \mb v_j ||_F^2 \nonumber \\
& \mathrm{s.t.}~~ ||\mb v_j ||_0 \leq T,~~j = 1, \cdots, J,
\label{opt:sc}
\end{align}
each of which can be efficiently solved by ``off-the-shelf'' sparsity-aware optimization algorithms \cite{theodoridis2020machine}, including orthogonal matching pursuit (OMP) \cite{tropp2007signal}, approximate message passing (AMP)\cite{donoho2010message}, and so forth. Then, given the learnt coefficient matrix $\mb V = [\mb v_1, \cdots, \mb v_J]$, K-SVD algorithm utilizes the K-means method for vector quantization (VQ) codebook design to give an updated dictionary matrix $\mb Q^{t+1}$.    {\color{black} The  iterative K-SVD algorithm  was shown to converge} to a local minima of the problem in Eq.~\eqref{opt:dl} \cite{aharon2006k}.

After the convergence of the K-SVD algorithm, the learnt basis functions for SSF representation are the columns of dictionary matrix $\mb Q$, which demands a large memory to maintain $I\times Z$ dictionary entries. For an unseen zero-mean SSF sample $\mb X^* \in \mathbb R^{I \times J^\prime}$, the coefficient matrix $\mb V^* = [\mb v_1^*, \cdots, \mb v_{J^\prime}^*] \in \mathbb R^{Z \times J^\prime}$ has no closed-form solution.  Instead, an iterative algorithm, e.g., OMP \cite{tropp2007signal}, needs to be resorted to solve a sparse coding problem (see problem \eqref{opt:sc}) for estimating each column $\mb v_{j^\prime}^*, \forall j^\prime $, which costs more computational resources than the EOF-based counterpart. Nevertheless,  numerical results have demonstrated that the basis functions learnt from K-SVD algorithm can improve the reconstruction performance of SSF \cite{bianco2017dictionary}, compared to the results using EOFs.

\subsection{3D SSF: 2D Fourier Basis Functions and 1D EOFs}
\label{sec:ii-c}

For a 3D SSF, as illustrated in Fig.~\ref{fig:FIG1}, one could first unfold it into a 2D SSF matrix and then apply the EOF-based\cite{leblanc1980underwater} or K-SVD-based\cite{aharon2006k, bianco2017dictionary} algorithm. The unfolding step, however, has broken the inherent 3D structure of SSF data, thereby leading to performance degradation.  {\color{black}A} classical method for 3D SSF representation relies on 2D Fourier basis functions and 1D EOFs \cite{cornuelle1989ocean, morawitz1996three}. The key idea is to use EOFs to capture the variations of SSF across different depths  {\color{black} and} use 2D Fourier basis functions to describe the horizontal slices of SSF. Specifically,
{\color{black} each de-mean SSF data in $\{ c(x,y,z)\}_{x=1,y=1,z=1}^{M,N,I}$ is }assumed to have the following expression\cite{cornuelle1989ocean, morawitz1996three}:
\begin{align}
&c (x,y,z) =  \sum_{f_1=1}^{N_{F_1}}\sum_{f_2=1}^{N_{F_2}} \sum_{k=1}^{{\color{black}K_F}}  w_{f_1,f_2,k}  \underbrace{[\mb E_{K_F}]_{z,k}}_{\text{1D EOF}} \nonumber\\
& \times \underbrace{ \exp \left(2 \pi j \left[ \frac{x(f_1 -1)}{L_x} \right] \right)  \exp \left(2 \pi j \left[ \frac{y(f_2 -1)}{L_y} \right] \right)}_{\text{2D Fourier basis function}},
\label{fourier}
\end{align}
where $[\mb E_{K_F}]_{z,k}$ denotes the $(z,k)$-th element of EOF matrix $\mb E_{K_F}$, which has $K_F$ leading-order EOFs. 
{\color{black}$w_{f_1,f_2,k}$ is the corresponding coefficient. $M$ and $N$ denote the two horizontal dimensions (i.e., length and width) of 3D SSF data.} $N_{F_1}$ and $N_{F_2}$ denote the number of Fourier basis functions for the two horizontal axes. $L_x$ and $L_y$ describe the periodicity of the associated Fourier basis function.

The EOF matrix $\mb E_{K_F}$ is obtained by firstly unfolding the 3D SSF along the vertical axis (as illustrated in Fig.~\ref{fig:FIG1}) and then  {\color{black} performing} the EVD on the resulting SSF matrix $\mb X^{\text{u}} \in \mathbb R^{I \times MN}$ (as introduced in Section \ref{Sec II-A}). Note that the matrix $\mb X^{\text{u}}$ describes the depth-range characteristics of 3D SSF. 
The 2D Fourier basis functions can be generated according to  {\color{black}Eq.~\eqref{fourier}}. For the brevity of notation, we define the  2D Fourier matrix $\mb F \in \mathbb C^{MN \times N_{F_1}  N_{F_2}}$ as follows:
\begin{align}
    \mb F = \mb F_2 \otimes  \mb F_1, 
\end{align}
where $\mb F_1$ and $\mb F_2$ are defined in  {\color{black} Eq.~(9) and (10)}, respectively:
\begin{align}
&\mb F_1 = \begin{bmatrix}
1 & 1 & \cdots & 1 \\
1 & \exp \left(\frac{2 \pi j }{L_x} \right)  & \cdots & \exp \left(2 \pi j \left[ \frac{N_{F_1} -1}{L_x} \right] \right)  \\
1& \exp \left( \frac{4 \pi j }{L_x} \right) & \cdots & \exp \left(2 \pi j \left[ \frac{2(N_{F_1} -1)}{L_x} \right] \right)\\
\vdots &  \vdots & \vdots & \vdots \\
1 & \exp \left( \frac{2(M-1)\pi j}{L_x}  \right) & \cdots & \exp \left(2 \pi j \left[ \frac{(M-1)(N_{F_1} -1)}{L_x} \right] \right)
\end{bmatrix} \nonumber \\
&~~~~~~~~~ \in \mathbb C^{M \times N_{F_1}}, 
\end{align}
\begin{align}
&\mb F_2 = \begin{bmatrix}
1 & 1 & \cdots & 1 \\
1 & \exp \left(\frac{2 \pi j }{L_y} \right)  & \cdots & \exp \left(2 \pi j \left[ \frac{N_{F_2} -1}{L_y} \right] \right)  \\
1& \exp \left( \frac{4 \pi j }{L_y} \right) & \cdots & \exp \left(2 \pi j \left[ \frac{2(N_{F_2} -1)}{L_y} \right] \right)\\
\vdots &  \vdots & \vdots & \vdots \\
1 & \exp \left( \frac{2(N-1)\pi j}{L_y}  \right) & \cdots & \exp \left(2 \pi j \left[ \frac{(N-1)(N_{F_2} -1)}{L_y} \right] \right)
\end{bmatrix} \nonumber \\
&~~~~~~~~~  \in \mathbb C^{N \times N_{F_2}}. 
\end{align}
Each column in $\mb F$ represents a 2D Fourier basis function.

Given the 1D EOFs and 2D Fourier basis functions, following  {\color{black}Eq.~\eqref{fourier}}, any unseen  3D SSF data $\{ c^*(x,y,z)\}_{x=1,y=1,z=1}^{M,N,I}$ can be represented by 3D coefficients $\{ w^*_{f_1,f_2,k} \}_{f_1=1,f_2=1,k=1}^{N_{F_1},N_{F_2},K_F}$. To efficiently compute these coefficients, both {\color{black} the 3D SSF data  $\{ c^*(x,y,z)\}_{x=1,y=1,z=1}^{M,N,I}$ and coefficients $\{ w^*_{f_1,f_2,k} \}_{f_1=1,f_2=1,k=1}^{N_{F_1},N_{F_2},K_F}$} can be unfolded into 2D matrices $\mb X^{*, \text{u}} \in \mathbb R^{I \times MN}$ and $\mb W^* \in \mathbb R^{K_F \times N_{F_1} N_{F_2}}$ {\color{black},} respectively (see Fig.~\ref{fig:FIG1}). {\color{black} The coefficients can be computed by} 
\begin{align}
    \mb W^* =  \mb E_{K_F}^\T \mb X^{*, \text{u}} (\mb F^\T)^{\dagger} {\color{black}.} 
\end{align}
Note that under this scheme, the number of coefficients required for SSF representation is $N_{F_1} N_{F_2} K_F $. 

{\color{black} {\it \textbf{Remark 1:}}  If the considered spatial area is large and the associated sound speeds have significant variations,  larger values of $N_{F_1}$ and $N_{F_2}$  should  be chosen. Otherwise, smaller values of $N_{F_1}$ and $N_{F_2}$  could be selected. If history data is available, trial-and-error method is viable to select these two values \cite{theodoridis2020machine}. In recent machine learning, Bayesian approach was leveraged to achieve automatic model order selection \cite{theodoridis2020machine, cheng2020learning,  zhao2015bayesian,xule21}. Namely, the hyper-parameters (e.g., $N_{F_1}$ and $N_{F_2}$) might be learnt directly from training data, which is an interesting future research direction.}

\section{Representation Learning Via Tensors}
\label{sec:3}
 Before moving to the exploration of more effective 3D SSF representation, we first provide some touches on the preliminaries of tensors\cite{kolda2009tensor}, including terminologies, tensor operations, and tensor decomposition formats. Then, we interpret tensor decomposition in the context of representation learning\cite{panagakis2021tensor, sidiropoulos2017tensor}, showing its paramount role in modern data science and ocean signal processing. 

\subsection{Scalar, Vector, Matrix, and Tensor}

\begin{figure}[h]
\includegraphics[width=1\reprintcolumnwidth]{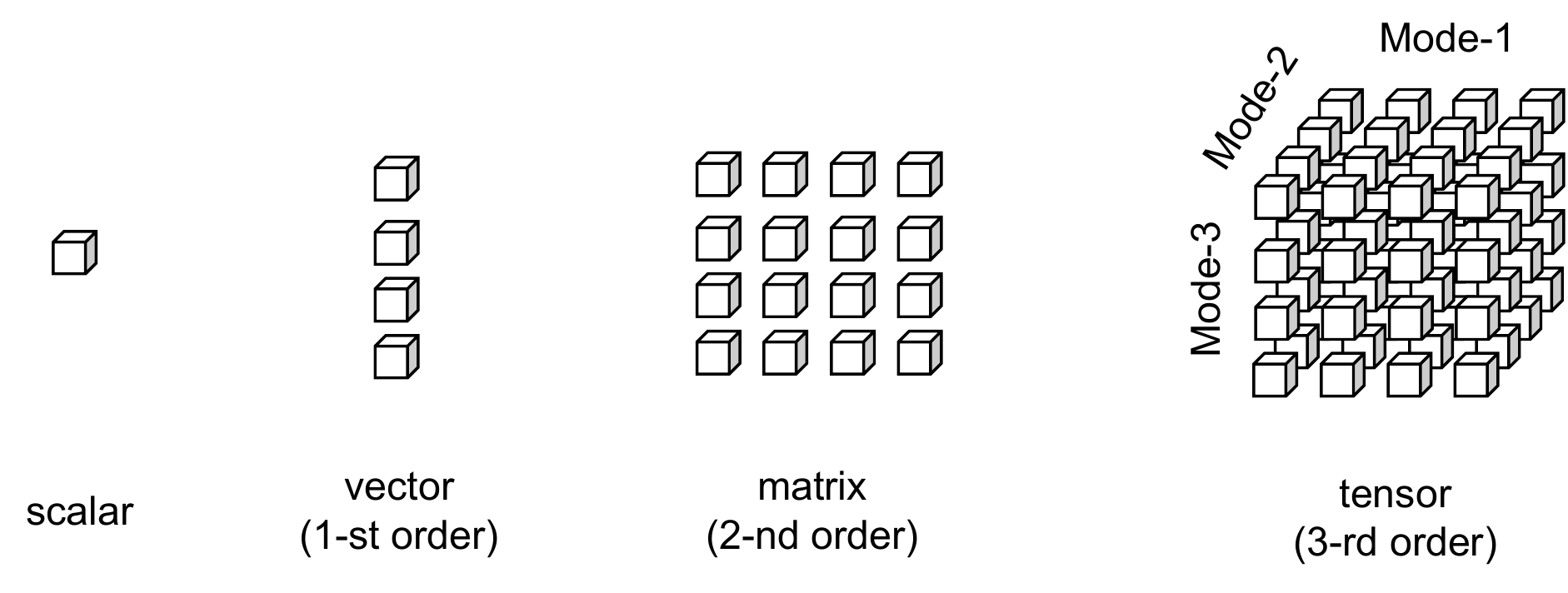}
\caption{\label{fig:FIG2}{ Illustration of scalar, matrix and tensor.}}
\end{figure}

In multilinear algebra, the term {\it order} measures the number of indices used to assess each data element (in scalar form)\cite{kolda2009tensor}. Specifically, vector $\mb a \in \mathbb C^{J_1}$ is the $1$-st order tensor since its element $\mb a_{j_1}$ can be assessed via only one index. Matrix $\mb A  \in \mathbb C^{J_1 \times J_2}$ is the $2$-nd order tensor, because two indices are enough to traverse all of its elements $\mb A_{j_1,j_2}$. As a generalization, tensors are of order three or higher. A $P$-th order tensor $\bc A \in \mathbb C^{J_1 \times \cdots \times J_P}$ utilizes $P$ indices to address its elements $\bc A_{j_1, \cdots, j_P}$. For illustration, we depict the scalar, vector, matrix and tensor in Fig.~\ref{fig:FIG2}. 

For a $P$-th order tensor $\bc A$, each index corresponds to a {\it mode}\cite{kolda2009tensor}, which is used to generalize the concepts of rows and columns of matrices to tensors. For example, for a third order tensor $\bc A \in \mathbb C^{J_1 \times J_2 \times J_3}$,  given indices $j_2$ and $j_3$, the vectors $\bc A_{:,j_2, j_3}$ are termed as {\it mode-1 fibers}.

{\it \textbf{Remark 2:}} The 3D SSF data $\{c(x,y,z)\}_{x=1,y=1,z=1}^{M,N,I}$ can be naturally represented by a third-order tensor $\bc X \in \mathbb R^{M \times N \times I}$, with each element $\bc X_{x,y,z}$ being  $c(x,y,z)$.

\subsection{Tensor Unfolding}

\begin{figure}[t]
\includegraphics[width=1\reprintcolumnwidth]{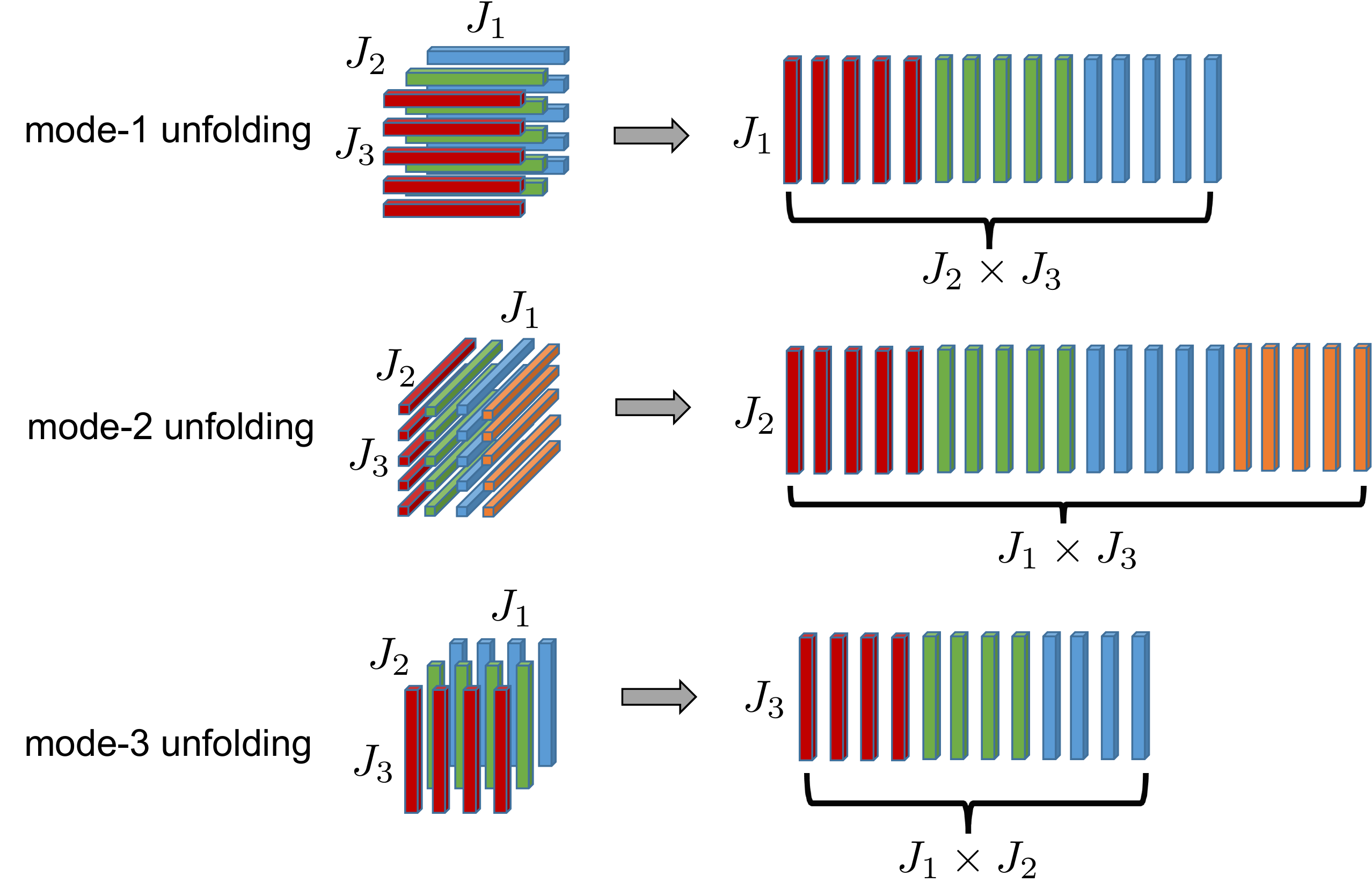}
\caption{\label{fig:FIG3}{ Illustration of tensor unfolding.}}
\end{figure}

Tensor unfolding aims to re-organize the fibers in one mode into a matrix. For a $P$-th order tensor $\bc A \in \mathbb C^{J_1 \times \cdots \times J_P}$, since it has $P$ modes, there are $P$ types of unfolding, each termed as {\it mode-$p$ unfolding}. {\color{black} It is formally defined as follows \cite{kolda2009tensor} and illustrated} in Fig.~\ref{fig:FIG3}.

\begin{tcolorbox}
\textbf{Definition 1 (Mode-$p$ Unfolding)} Given a tensor $\bc A \in \mathbb C^{J_1 \times \cdots \times J_P}$, its mode-$p$ unfolding gives a matrix $\mb A_{(p)} \in \mathbb C^{J_p \times \prod_{k=1, k \neq p}^P  J_k}$. Each tensor element $\bc A_{j_1,\cdots,j_P}$ is mapped to the matrix element $ \left [\mb A_{(p)}\right]_{j_p,q}$, where $q = 1 + \sum_{k=1, k\neq p}^P (j_k -1) I_k$ with $I_k = \prod_{m = 1, m\neq p }^{k-1} J_m$.
\end{tcolorbox}

Tensor unfolding is one of the most important operations in tensor-based machine learning and signal processing\cite{panagakis2021tensor, sidiropoulos2017tensor}, since it gives a ``matrix'' view to describe a tensor data, such that fruitful results in linear algebra can be leveraged. {\color{black} Typically}, tensor-based  algorithms were mostly developed upon the matrices provided by unfolding operations.

Then,  the $p$-mode product between a tensor  and a matrix  {\color{black} is introduced} as follows \cite{kolda2009tensor} . 

\begin{tcolorbox}
\textbf{Definition 2 ($p$-mode Product)} The $p$-mode product between a tensor $\bc A \in \mathbb C^{J_1 \times \cdots \times J_P}$ and a matrix $\mb M \in \mathbb C^{R \times J_p}$ results in a tensor $(\bc A \times_{p} \mb M)$ $\in \mathbb C^{J_1 \times \cdots \times  J_{p-1} \times R \times J_{p+1} \times \cdots \times J_P}$, with each element being
\begin{align}
&(\bc A \times_{p} \mb M )_{j_1,\cdots, j_{p-1},r,j_{p+1},\cdots, j_P} \nonumber \\
& = \sum_{j_p =1}^{J_P}  m_{r,j_p} \bc A_{j_1,\cdots, j_P}.
\end{align}
\end{tcolorbox}

{\it \textbf{Remark 3:}} The unfolding rule introduced in Fig.~\ref{fig:FIG1} is essentially the mode-$3$ unfolding of a 3D tensor. That is, the unfolding matrix $\mb X^{\text{u}}$ in Section~\ref{sec:ii-c} corresponds to $\mb X_{(3)}$ introduced in Definition 1.

\subsection{Tensor Decomposition for Representation Learning}
\label{sec:iii-c}
To extract low-dimensional yet informative parameters (in terms of smaller tensors, matrices and vectors) from multi-dimensional data, tensor decomposition, which generalizes matrix decomposition to tackle higher-order tensors, has come up as the major tool in recent machine learning and signal processing studies \cite{panagakis2021tensor, sidiropoulos2017tensor}. {\color{black} The extracted parameters are expected to preserve the structures endowed by physical sciences and have clear interpretations.} To achieve this goal, various tensor decomposition formats  \cite{kolda2009tensor} have been proposed, in which canonical polyadic decomposition (CPD) and Tucker decomposition are the most well-known and widely adopted. 

In this paper, we focus on tensor Tucker decomposition, {\color{black} which} includes CPD as a special case. {\color{black} The definition of Tucker decomposition} is given as follows  \cite{kolda2009tensor}.

\begin{figure}
\includegraphics[width=1\reprintcolumnwidth]{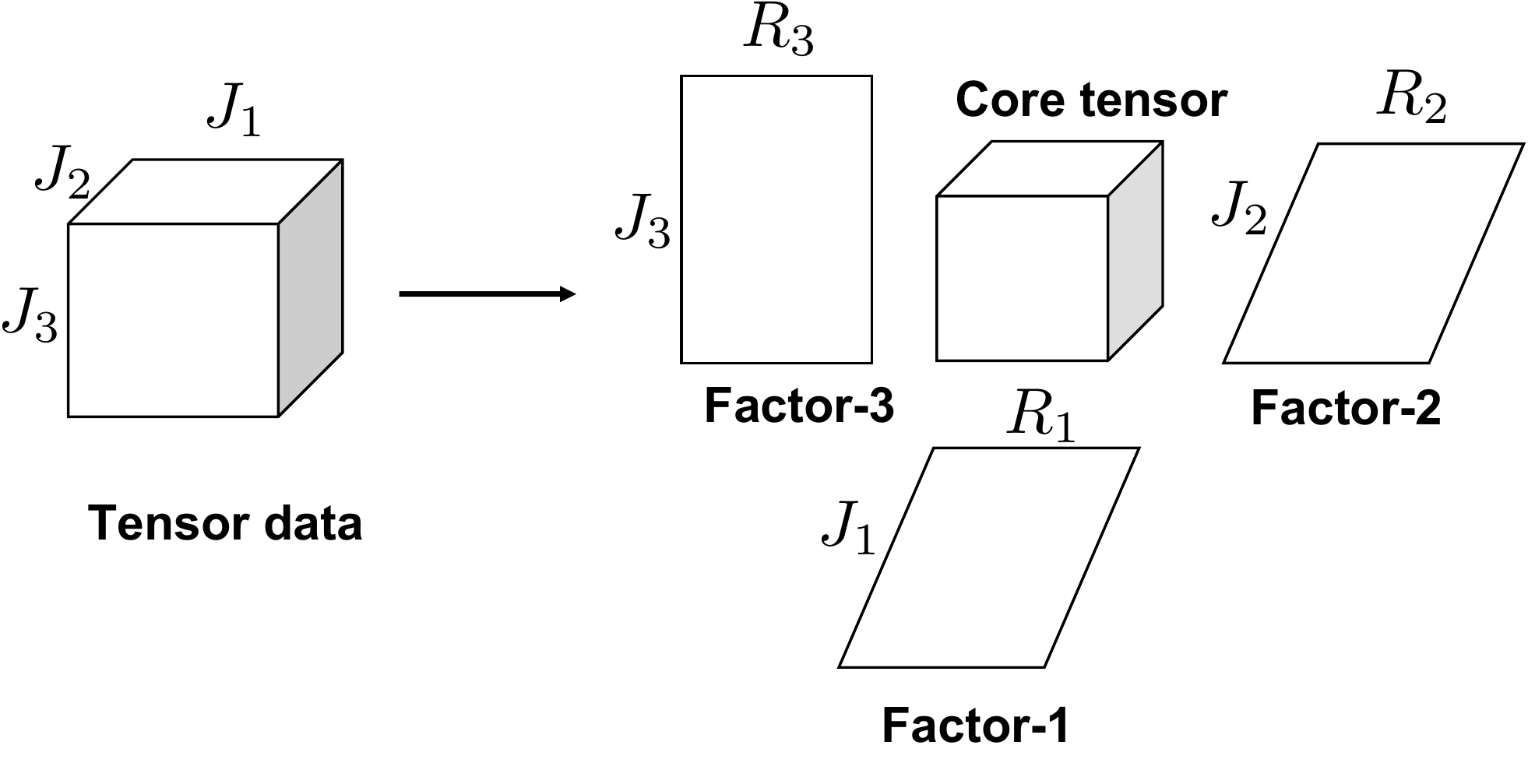}
\caption{\label{fig:FIG4}{ Illustration of tensor Tucker decomposiiton.}}
\end{figure}

\begin{tcolorbox}
\textbf{Definition 3 (Tucker Decomposition)}
 For a $P$-th order tensor $\bc A \in \mathbb C^{J_1 \times \cdots \times J_P}$, tensor Tucker decomposition is defined as
\begin{align}
\bc A = \bc G \times_1 \mb U^{(1)} \times_2 \mb U^{(2)}  \times_3 \cdots \times_P \mb U^{(P)},
\label{tuckerd}
\end{align}
where each factor matrix $\mb U^{(p)} \in \mathbb C^{J_p \times R_p}, {\color{black} \forall p = 1,2,  \cdots, P}$, and is usually orthonormal. The core tensor {\color{black} is} $\bc G \in \mathbb C^{R_1 \times R_2 \times \cdots \times R_P}$. The tuple $(R_1, \cdots, R_P)$ is known as {\color{black} the} multi-linear rank. 
\end{tcolorbox}
Note that the definition above utilizes the  $p$-mode product (see Definition 2). The illustration of tensor Tucker decomposition is provided in Fig.~\ref{fig:FIG4}. Usually, we have  $R_p \ll J_p, \forall p$. Note that when the core tensor $\bc G$ is super-diagonal and $R_1 = \cdots = R_P$, Tucker decomposition reduces to CPD  \cite{kolda2009tensor} .  

More insights can be drawn after interpreting Tucker decomposition \eqref{tuckerd} in the context of representation learning. In particular, the factor matrices  $\{\mb U^{(p)}  \}_{p=1}^P$ can be treated as the dictionary matrices, thereby providing a common set of basis functions for data representation. On the other hand, the core tensor $\bc G$, as seen in  \eqref{tuckerd}, acts as the weighting coefficients that encode the information of  
tensor data. In other words, relying on Tucker decomposition, the essence  of tensor-based representation learning is to acquire the factor matrices $\{\mb U^{(p)}  \}_{p=1}^P$ (i.e., basis functions) from training data $\bc A$, based on which any unseen/test data $\bc A^*$ can be represented by the core tensor $\bc G^*$. We make this interpretation concretely using the 3D SSF data in the next section.

\section{Tensor-based Basis Function Learning}
\label{sec:4}

In this section, 3D SSF representation {\color{black} is re-visited}  under the lens of tensor decomposition. In contrast to Section~\ref{sec:ii-c} that designs the basis functions in an empirical manner,  we view the 3D SSF data  as a third-order tensor $\bc X \in \mathbb R^{M \times N \times I}$(see Remark 2), and propose to learn the basis functions via a data-driven approach. Then, theoretical insights are given that interpret the classical basis functions (using EOFs and/or  Fourier basis functions)\cite{leblanc1980underwater, cornuelle1989ocean, morawitz1996three} as the special cases of the proposed tensor-based learning framework. {\color{black} Finally, if multiple 3D SSFs $\{\bc X_t \in \mathbb R^{M \times N \times I}\}_{t=1}^T$ (e.g., from different seasons) are available as training data,  we extend the proposed tensor-based basis function learning to jointly process these 3D SSFs.}

\subsection{Tensor-based Basis Function Learning Framework}
\label{sec:HOOI}
{\color{black} In this subsection,  a tensor-based basis function learning framework relying on Tucker decomposition is introduced, under which the higher-order orthogonal iteration (HOOI) algorithm is presented to learn the basis functions from one 3D SSF $\bc X \in \mathbb R^{M \times N \times I}$. As introduced in Section~\ref{sec:iii-c},  basis functions are provided by the three factor matrices in Eq.~\eqref{tuckerd}. In the context of representing 3D SSF  $\bc X$, they are denoted by $ \mb B^{(1)} \in \mathbb R^{M \times L_1} $, $ \mb B^{(2)} \in \mathbb R^{N \times L_2}$ and $ \mb B^{(3)} \in \mathbb R^{I \times L_3}$, respectively. The core tensor, which contains the coefficients for SSF representation, is denoted by  $\bc S \in \mathbb R^{L_1 \times L_2 \times L_3}$. Consequently, we propose the tensor-based basis function learning framework as follows:}
\begin{align}
    &\min_{\bc S, \mb B^{(1)}, \mb B^{(2)},\mb B^{(3)}} \left \| \bc X -  \bc S \times_1 \mb B^{(1)} \times_2 \mb B^{(2)}  \times_3   \mb B^{(3)}  \right\|_\F^2, \nonumber\\
    & \text{s.t.}~~  \bc S \in \mathbb R^{L_1 \times L_2 \times L_3}, \nonumber \\
    &~~~~~ f(\bc S) \geq  \mb 0, ~ \Bar{f} (\bc S) =  \mb 0, \nonumber \\
    & ~~~~~ \mb B^{(1)} \in \mathbb R^{M \times L_1}, ~ \mb B^{(2)} \in \mathbb R^{N \times L_2}, ~ \mb B^{(3)} \in \mathbb R^{I \times L_3}, \nonumber \\ 
    &~~~~~ g_p(\mb B^{(p)}) \geq \mb 0,  ~ \Bar{g}_p(\mb B^{(p)}) = \mb 0, ~ p = 1,2,3,
    \label{eq14}
\end{align}
where $f(\cdot)$ and $g_p(\cdot)$ denote the inequality constraints of the argument; and $\Bar{f}(\cdot)$ and $\Bar{g}_p(\cdot)$  represent the equality constraints of the argument. We can devise these constraints by incorporating the prior  knowledge of SSF,  or for the saving of computational resources.  {\color{black} For example,  practitioners can devise these constraint functions to embed the structures (e.g., non-negativeness, orthogonality, smoothness) into the basis function learning \cite{sidiropoulos2017tensor, panagakis2021tensor, cheng2020learning}. }

\begin{table}[!t]
\begin{tcolorbox}
{\textbf{\color{black} Algorithm 1:  (The HOOI Algorithm)}}
\\
\centering
\begin{ruledtabular}
\begin{tabular}{c} 
\leftline{\textbf {Input:} $\bc X \in \mathbb R^{M \times N \times I}$, multi-linear rank $(L_1, L_2, L_3)$.}\\
\leftline{\textbf{Initialize:}  $\mb B^{(1),0},  \mb B^{(2),0},  \mb B^{(3),0}$.} \\
\leftline{\textbf{For} $t = 1,2,3,\cdots$} \\
\leftline{ ~~~~$\mb C_{(1)}^{t} = \mb X_{(1)} ( \mb B^{(3),t-1} \otimes \mb B^{(2),t-1})$,}  \\ 
\leftline{  ~~~~$\mb B^{(1),t} \leftarrow$ $L_1$ leading left singular vectors of $\mb C_{(1)}^{t}$,}  \\ 
\leftline{ ~~~~$\mb C_{(2)}^{t} = \mb X_{(2)} ( \mb B^{(3),t-1} \otimes \mb B^{(1),t})$,} \\ 
\leftline{ ~~~~$\mb B^{(2),t} \leftarrow$ $L_2$ leading left singular vectors of $\mb C_{(2)}^{t}$,}  \\ 
\leftline{ ~~~~$\mb C_{(3)}^{t} = \mb X_{(3)} ( \mb B^{(2),t-1} \otimes \mb B^{(1),t})$,}  \\ 
\leftline{ ~~~~$\mb B^{(3),t} \leftarrow$ $L_3$ leading left singular vectors of $\mb C_{(3)}^{t}$,} \\ 
\leftline{\textbf{Until Convergence}} \\
\leftline{ $\bc S^{t} = \bc X \times_1 [\mb B^{(1),t}]^\T \times_2 [\mb B^{(2),t}]^\T  \times_3   [\mb B^{(3),t}]^\T$.} \\
\leftline{\textbf{Return} $\bc S^{t},  \mb B^{(1),t}, \mb B^{(2),t}, \mb B^{(3),t}.$
}
\end{tabular}
\end{ruledtabular}
\end{tcolorbox}
\label{tab1}
\end{table}

To reduce the number of matrix inversion, which is computationally demanding, orthonormal constraints are usually imposed on the factor matrices. {\color{black} More specifically, $\bar{g}_p(\mb B^{(p)}) = \mb 0$ is set to be:  $\left[ \mb B^{(p)} \right]^T   \mb B^{(p)}  - \mb I  = \mb 0$ , i.e., $\left[ \mb B^{(p)} \right]^T   \mb B^{(p)}  = \mb I $, where $\mb I$ is the identity matrix with matching dimensions.  Besides this, other constraints are not devised. Then the problem in Eq. \eqref{eq14}  reduces to the following problem:}
\begin{align}
    &\min_{\bc S, \mb B^{(1)}, \mb B^{(2)},\mb B^{(3)}} \left \| \bc X -  \bc S \times_1 \mb B^{(1)} \times_2 \mb B^{(2)}  \times_3   \mb B^{(3)}  \right \|_{\F}^2, \nonumber\\
    & \text{s.t.} ~~ \bc S \in \mathbb R^{L_1 \times L_2 \times L_3}, \nonumber \\
    & ~~~~~ \mb B^{(1)} \in \mathbb R^{M \times L_1}, ~[\mb B^{(1)}]^\T \mb B^{(1)} = \mb I_{L_1} \nonumber \\
    & ~~~~~ \mb B^{(2)} \in \mathbb R^{N \times L_2}, ~[\mb B^{(2)}]^\T \mb B^{(2)} = \mb I_{L_2} \nonumber \\
    & ~~~~~ \mb B^{(3)} \in \mathbb R^{I \times L_3}, ~[\mb B^{(3)}]^\T \mb B^{(3)} = \mb I_{L_3}. 
    \label{eq15}
\end{align}
If factor matrices $\{\mb B^{(p)}\}_{p=1}^3$ are given, the core tensor $\bc S$ can be solved to be \cite{kolda2009tensor}
\begin{align}
    \bc S = \bc X \times_1 [\mb B^{(1)}]^\T \times_2 [\mb B^{(2)}]^\T  \times_3   [\mb B^{(3)}]^\T.
    \label{eq17}
\end{align}
After substituting {\color{black} Eq.~\eqref{eq17}} into {\color{black} Eq.~\eqref{eq15}}, expanding the Frobenius norm and utilizing the orthonormal property of factor matrices, the problem in {\color{black} Eq.~\eqref{eq15}} is equivalent to the following problem:
\begin{align}
   & \max_{\{\mb B^{(p)}\}_{p=1}^3} \left \| \bc X \times_1 [\mb B^{(1)}]^\T \times_2 [\mb B^{(2)}]^\T  \times_3   [\mb B^{(3)}]^\T \right \|_{\F}^2, \nonumber \\
    &\text{s.t.} ~~ \mb B^{(1)} \in \mathbb R^{M \times L_1}, ~[\mb B^{(1)}]^\T \mb B^{(1)} = \mb I_{L_1} \nonumber \\
    & ~~~~~~ \mb B^{(2)} \in \mathbb R^{N \times L_2}, ~[\mb B^{(2)}]^\T \mb B^{(2)} = \mb I_{L_2} \nonumber \\
    & ~~~~~~ \mb B^{(3)} \in \mathbb R^{I \times L_3}, ~[\mb B^{(3)}]^\T \mb B^{(3)} = \mb I_{L_3}. 
    \label{eq18}
\end{align}
Problem \eqref{eq18} can be solved via alternating least-squares (ALS) method. In the $t$-th iteration, after fixing factor matrices $\{ \mb B^{(p),t-1}\}_{p = 2}^3$, problem \eqref{eq18} reduces to 
\begin{align}
    & \max_{\mb B^{(1)} \in \mathbb R^{M \times L_1}} \left  \| \left[\mb B^{(1)}\right]^\T \mb C_{(1)}^{t} \right \|_\F^2, \nonumber \\
       & \text{s.t.}  ~~ [\mb B^{(1)}]^\T \mb B^{(1)} = \mb I_{L_1}, 
       \label{eq181}
\end{align}
where 
\begin{align}
    \mb C_{(1)}^{t} = \mb X_{(1)} ( \mb B^{(3),t-1} \otimes \mb B^{(2),t-1}).
\end{align}
Note that $\mb X_{(1)}$ is the mode-$1$ unfolding matrix of tensor data $\bc  X$ (see Definition 1). {\color{black} The solution to the problem in Eq.~\eqref{eq181}} can be acquired via the singular value decomposition (SVD) of matrix $\mb C_{(1)}^{t}$, giving the following update step:
\begin{align}
   \mb B^{(1),t} = \left[ \mb u^{t}_1, \mb u^{t}_2, \cdots, \mb u^{t}_{L_1}\right],
\end{align}
where $\{ \mb u^{t}_l \}_{l=1}^{L_1}$ are $L_1$ leading left singular vectors of $\mb C_{(1)}^{t}$. Similar update steps can be derived for other two factor matrices $\{ \mb B^{(p),t}\}_{p = 2}^3$. Using these results, the algorithm that solves {\color{black} the problem  in Eq.~\eqref{eq15}} is summarized in Algorithm 1, which is known as {\it higher-order orthogonal iteration} (HOOI) algorithm \cite{kroonenberg1980principal}. {\color{black} The HOOI algorithm was proved to be convergence guaranteed \cite{kroonenberg1980principal}. }

Using the HOOI algorithm, basis functions (i.e., $\{\mb B^{(p)} \}_{p=1}^3$) can be learnt from training SSF data $\bc X$. For representing any unseen/test 3D SSF data $\bc X^*$, the core tensor $\bc S^*$, which has $L_1 L_2 L_3$ parameters, can be  learnt via {\color{black} Eq.~\eqref{eq17}}.

\subsection{Theoretical Insights}
\label{sec:iv-b}
{\color{black} A close connection exists between}  the proposed tensor-based basis function learning framework \eqref{eq14} and the classical basis functions using EOFs and Fourier basis functions\cite{leblanc1980underwater, cornuelle1989ocean, morawitz1996three} (as introduced in Section~\ref{Sec II-A} and Section~\ref{sec:ii-c} ). {\color{black} The connection is revealed  in the following two propositions, and the proofs are found in the Appendices. }

\begin{table}[!t]
\begin{ruledtabular}
\caption{\label{tab:prop}Differences among different basis functions under the unified tensor perspective.}
\begin{tabular}{|c|c|c|c|}
Basis functions &
  \begin{tabular}[c]{@{}c@{}}Factor-1\\  $\mb B^{(1)}$ \end{tabular} &
  \begin{tabular}[c]{@{}c@{}}Factor-2\\  $\mb B^{(2)}$ \end{tabular} &
  \begin{tabular}[c]{@{}c@{}}Factor-3\\   $\mb B^{(3)}$ \end{tabular} \\ \hline
EOFs  &
  \begin{tabular}[c]{@{}c@{}}Identity \\ matrix\end{tabular} &
  \begin{tabular}[c]{@{}c@{}}Identity\\  matrix\end{tabular} &
  \begin{tabular}[c]{@{}c@{}}{\bf Learnt} \\ {\bf from data}\end{tabular} \\ \hline
\begin{tabular}[c]{@{}c@{}}2D Fourier \\ basis functions\\ + 1D EOFs \end{tabular} &
  \begin{tabular}[c]{@{}c@{}}Fourier \\ matrix\end{tabular} &
  \begin{tabular}[c]{@{}c@{}}Fourier\\  matrix\end{tabular} &
  \begin{tabular}[c]{@{}c@{}}{\bf Learnt}\\  {\bf from data}\end{tabular} \\ \hline
HOOI-based &
  \begin{tabular}[c]{@{}c@{}}{\bf Learnt} \\ {\bf from data}\end{tabular} &
  \begin{tabular}[c]{@{}c@{}}{\bf Learnt} \\ {\bf from data}\end{tabular} &
  \begin{tabular}[c]{@{}c@{}}{\bf Learnt} \\ {\bf from data}\end{tabular} \\ 
\end{tabular}
\end{ruledtabular}
\end{table}

\begin{tcolorbox}
\textbf{Proposition 1.} The classical basis functions for 2D SSF,  expressed by the EOF matrix $\mb E_{K} \in \mathbb C^{I \times K}$, are the optimal solution of the following problem:
\begin{align}
    &\min_{\bc S, \mb B^{(1)}, \mb B^{(2)},\mb B^{(3)}} \left \Vert \bc X -  \bc S \times_1 \mb B^{(1)} \times_2 \mb B^{(2)}  \times_3   \mb B^{(3)} \right \Vert_\F^2, \nonumber\\
    & \text{s.t.} ~~ \bc S \in  \mathbb R^{M \times N \times K}, \nonumber \\
    & ~~~~~ \mb B^{(1)} = \mb I_{M} \in \mathbb R^{M \times M}, \nonumber \\
    & ~~~~~ \mb B^{(2)} = \mb I_{N} \in \mathbb R^{N \times N}, \nonumber \\
    & ~~~~~ \mb B^{(3)} \in \mathbb R^{I \times K}, ~[\mb B^{(3)}]^\T \mb B^{(3)} = \mb I_{K}, 
    \label{eq221}
\end{align}
which is a special case of the proposed tensor-based basis function learning framework \eqref{eq14}. \\

\noindent {\it Proof: See Appendix~\ref{appendix-a}.}
\end{tcolorbox}

\begin{tcolorbox}
\textbf{Proposition 2.} The classical basis functions for 3D SSF,  expressed by the EOF matrix $\mb E_{K_F} \in \mathbb C^{I \times K_F}$ and the two Fourier matrices $\mb F_1 \in \mathbb C^{M \times N_{F_1}}, \mb F_2 \in \mathbb C^{N \times N_{F_2}}$, are the optimal solution of the following problem:
\begin{align}
    &\min_{\bc S, \mb B^{(1)}, \mb B^{(2)},\mb B^{(3)}} \left \Vert \bc X -  \bc S \times_1 \mb B^{(1)} \times_2 \mb B^{(2)}  \times_3   \mb B^{(3)} \right \Vert_\F^2, \nonumber\\
    & \text{s.t.} ~~ \bc S \in  \mathbb C^{N_{F_1} \times N_{F_2} \times K_F}, \nonumber \\
    & ~~~~~ \mb B^{(1)} = \mb F_1 \in \mathbb C^{M \times N_{F_1}}, \nonumber \\
    & ~~~~~ \mb B^{(2)} = \mb F_2 \in \mathbb C^{N \times N_{F_2}}, \nonumber \\
    & ~~~~~ \mb B^{(3)} \in \mathbb R^{I \times K_F}, ~[\mb B^{(3)}]^\T \mb B^{(3)} = \mb I_{K_F}, 
    \label{eq22}
\end{align}
which is a special case of the proposed tensor-based basis function learning framework \eqref{eq14}. \\

\noindent {\it Proof: See Appendix~\ref{appendix-b}.}
\end{tcolorbox}

Proposition 1 and Proposition 2 point out that the classical basis functions  \cite{leblanc1980underwater, cornuelle1989ocean, morawitz1996three} and the one learnt using the HOOI algorithm are all special cases  of the proposed tensor-based basis function learning framework \eqref{eq14}. Through this unified perspective, the differences of these three types of basis functions are evident, as shown in Table~\ref{tab:prop}. Particularly, in problem \eqref{eq221},  {\color{black} the two factor matrices} are restricted to be the identity matrices, which are  too rigid to allow effective SSF representation. On the other hand,  in problem \eqref{eq22},  the two factor matrices are designed  to the Fourier basis matrices, thus with a higher representation capability than the identity matrices. The identity matrices and Fourier basis matrices are manually designed.  On the contrary, in problem \eqref{eq15}, three factor matrices are all learnt from the data. Therefore, problem \eqref{eq15} and the associated HOOI algorithm endue the basis functions (expressed by the learnt factor matrices) a higher flexibility, making them a promising candidate for more effective 3D SSF representation, as corroborated in Section~\ref{sec:5}.

\subsection{Learning Basis Functions  from Multiple 3D SSFs}
  
In Section~\ref{sec:HOOI} and \ref{sec:iv-b}, tensor-based basis function learning using one 3D SSF is introduced. Due to the relatively high correlations of 3D SSFs in several tens of days (e.g., one month)\cite{munk2009ocean, zhu2020}, the learnt basis functions are capable of reduced-order yet accurate 3D SSF representation in such a period. Therefore, using only one 3D SSF $\bc X \in \mathbb R^{M \times N \times I}$, the proposed approach is useful for underwater applications that require short-term SSF forecasting/inversion, e.g.,   geoacoustic inversion\cite{jiang2008short},  shallow water sound speed profile inversion\cite{zhang2015inversion}, internal wave reconstruction and acoustic propagation calculation\cite{casagrande2011novel}.

On the other hand, if multiple 3D SSFs  $\{\bc X_t \in \mathbb R^{M \times N \times I}\}_{t=1}^T$ (e.g., from different seasons) are available, the joint learning of basis functions has potential to realize long-term effective representation\cite{lu2004spatial, long2021variations}, since more SSF variations (e.g., from different seasons) are taken into account. To achieve this goal, we extend the tensor-based basis function learning problem in Eq.~\eqref{eq15} from dealing with one 3D SSF to processing multiple 3D SSFs:
\begin{align}
    &\min_{\bc S_t, \mb B^{(1)}, \mb B^{(2)},\mb B^{(3)}} \sum_{t=1}^T \left \| \bc X_t -  \bc S_t \times_1 \mb B^{(1)} \times_2 \mb B^{(2)}  \times_3   \mb B^{(3)}  \right \|_{\F}^2, \nonumber\\
    & \text{s.t.} ~~ \bc S_t \in \mathbb R^{L_1 \times L_2 \times L_3}, \nonumber \\
    & ~~~~~ \mb B^{(1)} \in \mathbb R^{M \times L_1}, ~[\mb B^{(1)}]^\T \mb B^{(1)} = \mb I_{L_1} \nonumber \\
    & ~~~~~ \mb B^{(2)} \in \mathbb R^{N \times L_2}, ~[\mb B^{(2)}]^\T \mb B^{(2)} = \mb I_{L_2} \nonumber \\
    & ~~~~~ \mb B^{(3)} \in \mathbb R^{I \times L_3}, ~[\mb B^{(3)}]^\T \mb B^{(3)} = \mb I_{L_3}. 
    \label{eq23}
\end{align}
Note that factor matrices $\{\mb B^{(p)} \}_{p=1}^3$ contain common basis functions for multiple 3D SSFs $\{ \bc X_t\}_{t=1}^T$,  and the core tensor $\bc S_t$ is associated with the 3D SSF $\bc X_t$, for $t = 1, \cdots, T$.

Using tensor algebra\cite{kolda2009tensor}, it can be shown that the problem in Eq.~\eqref{eq23} is equivalent to 
\begin{align}
    &\min_{ \tilde{\bc S}, \mb B^{(1)}, \mb B^{(2)},\mb B^{(3)}} \left \| \tilde{\bc  X} -   \tilde{\bc S} \times_1 \mb B^{(1)} \times_2 \mb B^{(2)}  \times_3   \mb B^{(3)}  \times_4 \mb I_{T}  \right \|_{\F}^2, \nonumber\\
    & \text{s.t.} ~~ \tilde{\bc S} \in \mathbb R^{L_1 \times L_2 \times L_3 \times T}, \nonumber \\
    & ~~~~~ \mb B^{(1)} \in \mathbb R^{M \times L_1}, ~[\mb B^{(1)}]^\T \mb B^{(1)} = \mb I_{L_1} \nonumber \\
    & ~~~~~ \mb B^{(2)} \in \mathbb R^{N \times L_2}, ~[\mb B^{(2)}]^\T \mb B^{(2)} = \mb I_{L_2} \nonumber \\
    & ~~~~~ \mb B^{(3)} \in \mathbb R^{I \times L_3}, ~[\mb B^{(3)}]^\T \mb B^{(3)} = \mb I_{L_3},
    \label{eq24}
\end{align}
where $\tilde{\bc  X} \in \mathbb R^{M\times N \times I \times T}$ and $\tilde{\bc S} \in \mathbb R^{L_1 \times L_2 \times L_3 \times T}$ are obtained by stacking $\{\bc X_t \in \mathbb R^{M \times N \times I}\}_{t=1}^T$  and $\{\bc S_t \in \mathbb R^{L_1\times L_2 \times L_3}\}_{t=1}^T$ along their fourth modes, respectively. 

According to Eq.~\eqref{tuckerd} in Definition 1, the problem in  Eq.~\eqref{eq24} is a variant of Tucker decomposition problem with one factor matrix being  an identity matrix $\mb I_T$. Therefore, a modified HOOI algorithm (labeled as M-HOOI)\cite{kolda2009tensor}, which is summarized in Algorithm 2, can be applied to solve the problem in  Eq.~\eqref{eq24}. The derivations of M-HOOI algorithm are similar to those presented in Section~\ref{sec:HOOI}.  

Using M-HOOI algorithm,  basis functions (i.e., $\{\mb B^{(p)} \}_{p=1}^3$) can be jointly learnt from multiple 3D SSFs $\{\bc X_t \}_{t=1}^T$. For representing an unseen 3D SSF $\bc X^*$, the coefficients in $\bc S^*$ can be acquired via Eq.~\eqref{eq17}.

\begin{table}[t]
\begin{tcolorbox}
{\textbf{\color{black} Algorithm 2:  (The M-HOOI Algorithm)}}
\\
\centering
\begin{ruledtabular}
\begin{tabular}{c} 
\leftline{\textbf {Input:} $\tilde {\bc X} \in \mathbb R^{M \times N \times I \times T}$, multi-linear rank $(L_1, L_2, L_3)$.}\\
\leftline{\textbf{Initialize:}  $\mb B^{(1),0},  \mb B^{(2),0},  \mb B^{(3),0}$.} \\
\leftline{\textbf{For} $t = 1,2,3,\cdots$} \\
\leftline{ ~~~~$\mb C_{(1)}^{t} = \mb X_{(1)} ( \mb I_T \otimes \mb B^{(3),t-1} \otimes \mb B^{(2),t-1})$,}  \\ 
\leftline{  ~~~~$\mb B^{(1),t} \leftarrow$ $L_1$ leading left singular vectors of $\mb C_{(1)}^{t}$,}  \\ 
\leftline{ ~~~~$\mb C_{(2)}^{t} = \mb X_{(2)} ( \mb I_T \otimes  \mb B^{(3),t-1} \otimes \mb B^{(1),t})$,} \\ 
\leftline{ ~~~~$\mb B^{(2),t} \leftarrow$ $L_2$ leading left singular vectors of $\mb C_{(2)}^{t}$,}  \\ 
\leftline{ ~~~~$\mb C_{(3)}^{t} = \mb X_{(3)} (\mb I_T \otimes  \mb B^{(2),t-1} \otimes \mb B^{(1),t})$,}  \\ 
\leftline{ ~~~~$\mb B^{(3),t} \leftarrow$ $L_3$ leading left singular vectors of $\mb C_{(3)}^{t}$,} \\ 
\leftline{\textbf{Until Convergence}} \\
\leftline{ $\tilde{\bc S}^{t} = \tilde{\bc X} \times_1 [\mb B^{(1),t}]^\T \times_2 [\mb B^{(2),t}]^\T  \times_3   [\mb B^{(3),t}]^\T  \times_4   \mb I_T $.} \\
\leftline{\textbf{Return} $\tilde{\bc S}^{t},  \mb B^{(1),t}, \mb B^{(2),t}, \mb B^{(3),t}.$
}
\end{tabular}
\end{ruledtabular}
\end{tcolorbox}
\label{tab1}
\end{table}

{\it \textbf{Remark 4:}} For the basis functions learnt via data-driven approaches, including EOF, K-SVD, and the  tensor-based methods, their performance will degrade when the test data become less correlated with training data. Consequently, the update of basis functions using new training data is required to maintain the good performance of basis functions. But fortunately, the processes driving ocean SSFs are inherently continuous in space and time, the update (or the re-training) of basis functions need not to be very frequent in most cases. There is a trade-off  between  the sustainability of learnt basis functions and the cost of training data over a span of time. If multiple 3D SSFs across a long span of time (e.g., different seasons) are used as the training data, the learnt basis functions are more likely to realize long-term effective SSF representation, as will be demonstrated in the next section.

{\it \textbf{Remark 5:}} The tensor-based basis function learning  leverages low-rank tensor decomposition  models to exploit  multi-dimensional  correlations inside 3D SSFs. It has advantages over classical matrix-based methods when the considered 3D SSFs can be  more accurately represented by low-rank tensor models. Under this perspective,  the requirements of 3D SSFs are introduced in Appendix~\ref{appendix-d}.

\section{Numerical results}
\label{sec:5}

In this section, numerical results are presented to showcase the excellent performance of the tensor-based basis function learning algorithms (i.e., HOOI algorithm and M-HOOI algorithm) for 3D SSF data representation.

\subsection{Learning Basis Functions from One 3D SSF}
\label{sec:v-a}

In this subsection, the performance of tensor-based basis functions learnt from one 3D SSF is evaluated. The training and test data, baseline algorithms and performance metrics adopted in this subsection are introduced as follows.

\begin{figure*}[!t]
\centering 
\includegraphics[width=1.5 \reprintcolumnwidth]{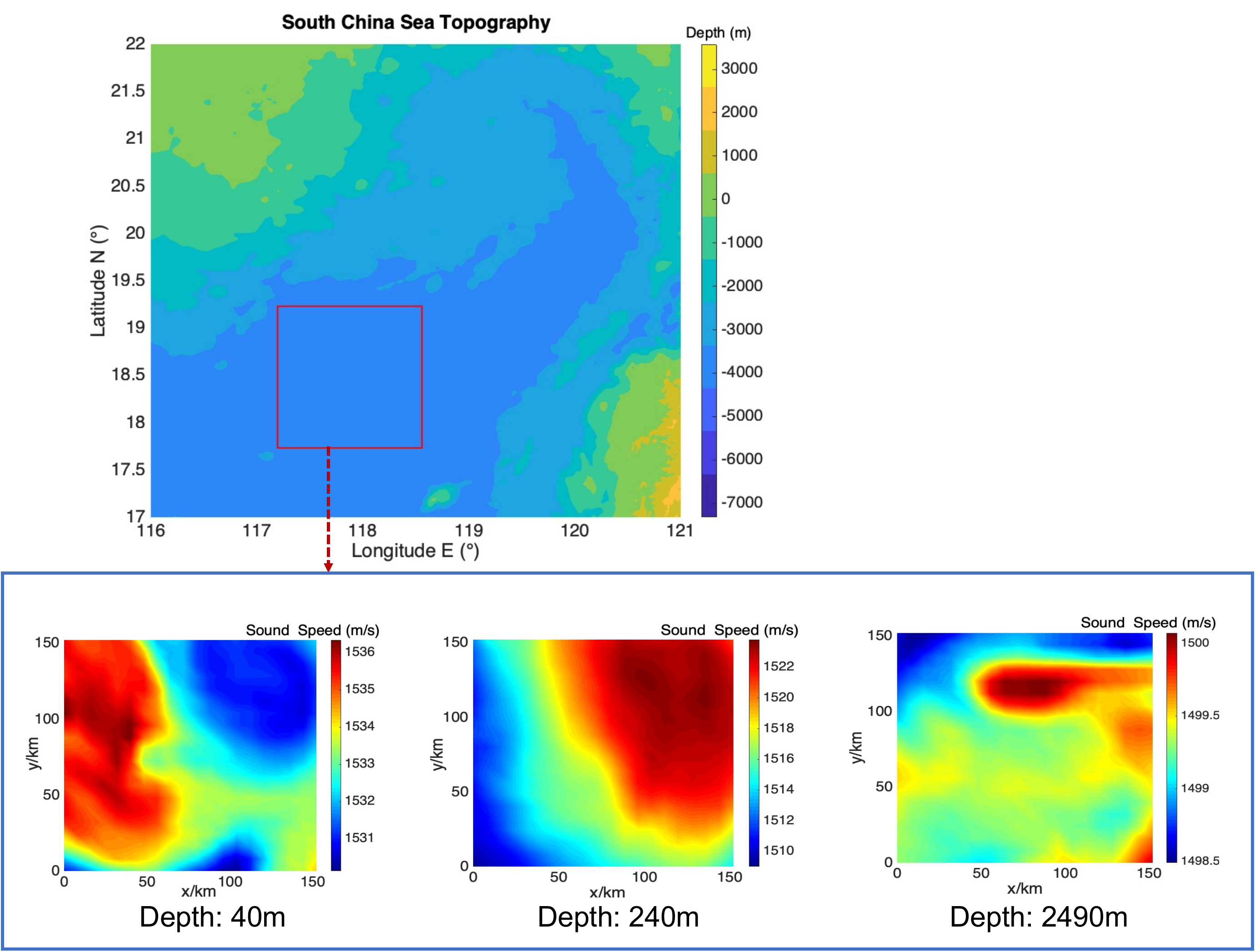}
\caption{Illustration of Training 3D SSF Data.}
\label{fig:FIGN2}
\hrule
\end{figure*}

\textbf{3D SSF Data:}  The 30-days 3D South China Sea (SCS)  SSF data $\{\bc X_t  \in \mathbb R^{20 \times 20 \times 300 }\}_{t=1}^{30}$ from Dec. 21, 2011 to Jan. 19, 2012 is analyzed in this paper. The data was derived by the 3D conductivity, temperature and depth (CTD) data across the area shown in Fig.~\ref{fig:FIGN2},  and was  provided by the Institute of Oceanology, Chinese Academy of Sciences using  a data-assimilative hybrid coordinate ocean model (HYCOM). We consider the 3D spatial area $152 \text{km} \times 152 \text{km} \times 2990 \text{m}$. That is, the horizontal resolution is {\color{black} $8 \text{km}$} and the vertical resolution is {\color{black} $10 \text{m}$}. For illustration, three horizontal slices of the $1$-st day SSF data $\bc X_1$,  corresponding to depths $40$m, $240$m, and $2490$m respectively,  are shown in Fig.~\ref{fig:FIGN2}.  In this area,  {\color{black} a mesoscale eddy can be observed}, which plays an important role in changing the ocean dynamics of a semi-closed ocean system\cite{zhu2020}.

\begin{figure*}[!t]
\baselineskip=12pt

\fig{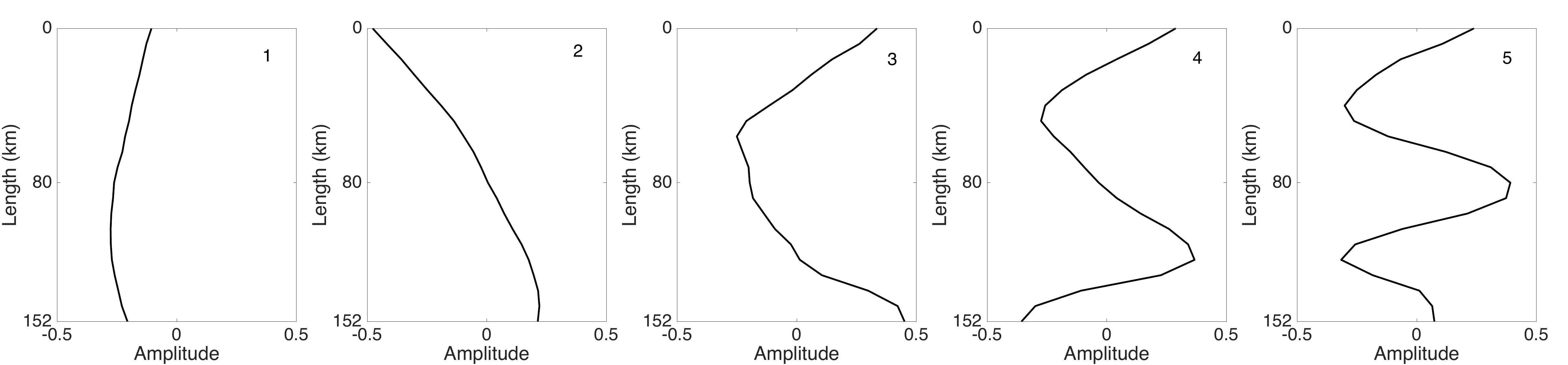}{2\reprintcolumnwidth}{(a)} \label{fig:FIGN1a}
\fig{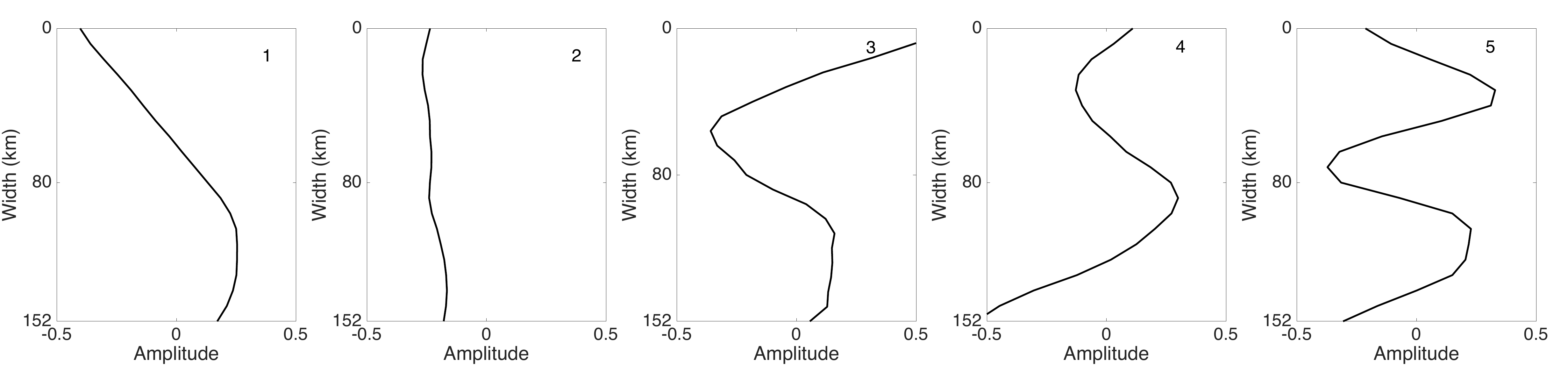}{2\reprintcolumnwidth}{(b)}  \label{fig:FIGN1b}
\fig{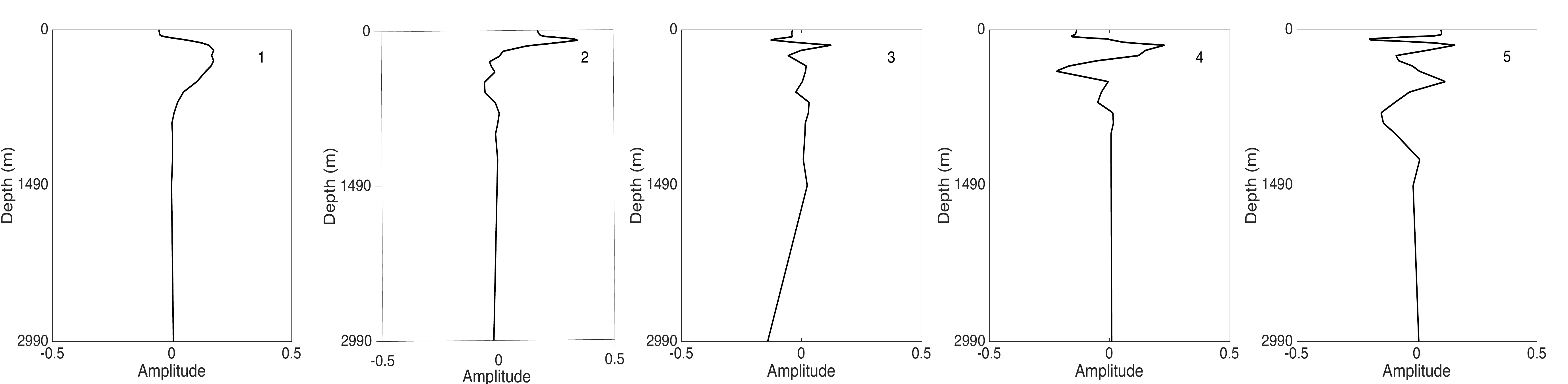}{2\reprintcolumnwidth}{(c)}  \label{fig:FIGN1c}
\caption{Illustration of tensor-based basis functions: (a) mode-1 basis functions, (b) mode-2 basis functions, and (c) mode-3 basis functions. }
\label{fig:FIGN1}
\hrule
\end{figure*}

\textbf{Training Data and Test Data:} The $1$-st day 3D SSF data{\color{black}, $\bc X_1$,} is used as the training data,  {\color{black} from which} the basis functions are learnt via the tensor-based HOOI algorithm and other benchmarking algorithms.  Visualization of the learnt tensor-based basis functions are provided in Fig.~\ref{fig:FIGN1}, in which the first $5$ columns of the three factor matrices $\{\mb B^{(p)} \}_{p=1}^3$ are plotted. From the mode-3 basis functions (expressed by the columns in  $\mb B^{(3)}$), it can be seen that the sound speeds vary much more significantly in the shallow ocean (the depth is smaller than $1000 \text{m}$), while change slightly in the deep ocean (the depth is larger than $2500 \text{m}$).  {\color{black} The basis functions expressed by the columns in $\mb B^{(1)}$ and $\mb B^{(2)}$ characterize the sound speed variations over the horizontal domain, which are not provided by classical matrix-based methods (e.g., EOFs).} The remaining 3D SSF data{\color{black}, $\{\bc X_t\}_{t=2}^{30}$,} are used as the test data to assess the representation capability of different basis functions. The data partition scheme follows the convention in ocean signal processing \cite{munk2009ocean,  zhu2020}.  {\color{black} Namely},  the $1$-st day SSF data is treated  as the history record and thus  {\color{black}  serves as} the training/reference.  {\color{black} Note that  the 3D spatial SSFs in 30 days are used to evaluate the performance of the algorithms, in order to see whether the basis functions learnt from a particular day are informative enough to represent the 3D SSFs of the following several tens of days (e.g., 29 days). In this regard, the time resolution for data partition is 1 day. The underlying assumption is that  the ocean sound speed variations can be well represented by the learnt basis functions  in at least 1-day period, which has been corroborated in the numerical study of this section.}

\textbf{Baselines:}  The benchmarking algorithms include the EOF-based method (labeled as EOF)\cite{leblanc1980underwater}, the K-SVD-based method (labeled as K-SVD)\cite{aharon2006k, bianco2017dictionary}, and the classical basis functions using 2D Fourier basis functions and 1D EOFs (labeled as 2D Fourier + 1D EOF)\cite{cornuelle1989ocean, morawitz1996three}.  In this paper, the K-SVD algorithm was implemented by the KSVD-Box v13 (\url{http://www.cs.technion.ac.il/~ronrubin/software.html}), where the OMP algorithm implemented by  OMP-Box v10 (\url{http://www.cs.technion.ac.il/~ronrubin/software.html}) was utilized for sparse coding.

\textbf{Performance Metrics:} The representation capabilities of different basis functions are assessed by the root mean square error (RMSE) of SSF reconstruction per horizontal slice,  defined by
\begin{align}
\text{RMSE}  = \frac{1}{I} \left \|  \bc X -  \hat{\bc X} \right \|_\F,
\end{align}
where $\hat{\bc X} $ is the reconstructed 3D SSF data; $\bc X$ is the ground-truth 3D SSF data; and $I = 300$ is the number of horizontal slices.  We also compare their running time in both training and testing process. All the experiments were conducted in Matlab R2019b with a 2.2 GHz 6-Core Intel Core i7 CPU.


\subsubsection{Reconstruction Error under The Similar Number of Representation Coefficients}

First, we assess the representation capability of different basis functions in terms of the reconstruction RMSEs under the similar number of representation coefficients. For the training data  $\bc X_1 \in \mathbb R^{20 \times 20 \times 300}$,  the tensor-based HOOI algorithm and other benchmarking algorithms (EOF, K-SVD, 2D Fourier + 1D EOF) were run to learn the corresponding basis functions, in which the EOF-based and K-SVD-based  {\color{black} algorithms} were performed on the  {\color{black} unfolded} 2D SSF matrix, as illustrated in Fig.~\ref{fig:FIG1}. The hyper-parameters of different algorithms were set to let their corresponding representation coefficients have similar numbers, as seen in Table~\ref{tab:table2}. {\color{black} Note that  the considered 3D SSF data is with $MN = 20\times 20 = 400$. Therefore, the number of coefficients for EOF and K-SVD are multiples of $400$, and cannot be an arbitrary number.   In Case I and Case II, two coefficient numbers (i.e., $800$ and $1200$) are considered for EOF and K-SVD scheme. Then we vary the hyper-parameters of other two algorithms to make their coefficient number closer to  $800$ and $1200$. Besides this, the reconstruction performance of different algorithms under a wide range of coefficient number is shown in Fig.~\ref{fig:FIGN10} and Fig.~\ref{fig:FIG7}.}

\begin{table*}
\caption{\label{tab:table2} The hyper-parameters and the number of representation coefficients for different algorithms. In each case, the number of representation coefficients for different algorithms is comparable.}

\begin{ruledtabular}
\begin{tabular}{|c|c|c|c|c|c|c|c|c|c|c|c|c|}
Cases                                                                                  & \multicolumn{6}{c|}{Case I} & \multicolumn{6}{c|}{Case II}        \\ \hline
Algorithms &
  HOOI &
  \multicolumn{2}{c|}{EOF} &
  \multicolumn{2}{c|}{K-SVD} &
  \begin{tabular}[c]{@{}c@{}}2D Fourier \\ + 1D EOF\end{tabular} &
  HOOI &
  \multicolumn{2}{c|}{EOF} &
  \multicolumn{2}{c|}{K-SVD} &
  \begin{tabular}[c]{@{}c@{}}2D Fourier\\ + 1D EOF\end{tabular} \\ \hline

  \begin{tabular}[c]{@{}c@{}}Hyper- \\ parameters \end{tabular}  &
  \begin{tabular}[c]{@{}c@{}}$L_1 =8$\\$L_2 =8$\\$L_3 = 10$\end{tabular} &
  $K=2$ &
  $K=3$ &
  \begin{tabular}[c]{@{}c@{}}$Z=320$\\ $T =2$ \end{tabular}  &
    \begin{tabular}[c]{@{}c@{}}$Z=320$\\ $T =3$ \end{tabular} &
  \begin{tabular}[c]{@{}c@{}}$N_{F_1} =8$\\ $N_{F_2} =8$\\ $K_F= 10$\end{tabular} &
  \begin{tabular}[c]{@{}c@{}}$L_1 =8$\\ $L_2 =10$\\ $L_3 = 10$\end{tabular} &
  $K = 2$ &
  $K=3$ &
  \begin{tabular}[c]{@{}c@{}}$Z=320$\\ $T =2$ \end{tabular}  &
  \begin{tabular}[c]{@{}c@{}}$Z=320$\\ $T =3$ \end{tabular} &
 \begin{tabular}[c]{@{}c@{}}$N_{F_1} =8$\\ $N_{F_2} =10$\\ $K_F= 10$\end{tabular} \\ \hline
\begin{tabular}[c]{@{}c@{}}The Number of \\ Representation\\ Coefficients\end{tabular} & 640  & 800 & 1200  & 800  &1200  & 640 & 800 & 800 &1200& 800 &1200 &800 \\ 
\end{tabular}
\end{ruledtabular}
\hrule
\end{table*}

\begin{figure*}[!t]
\baselineskip=12pt
\figline{
\fig{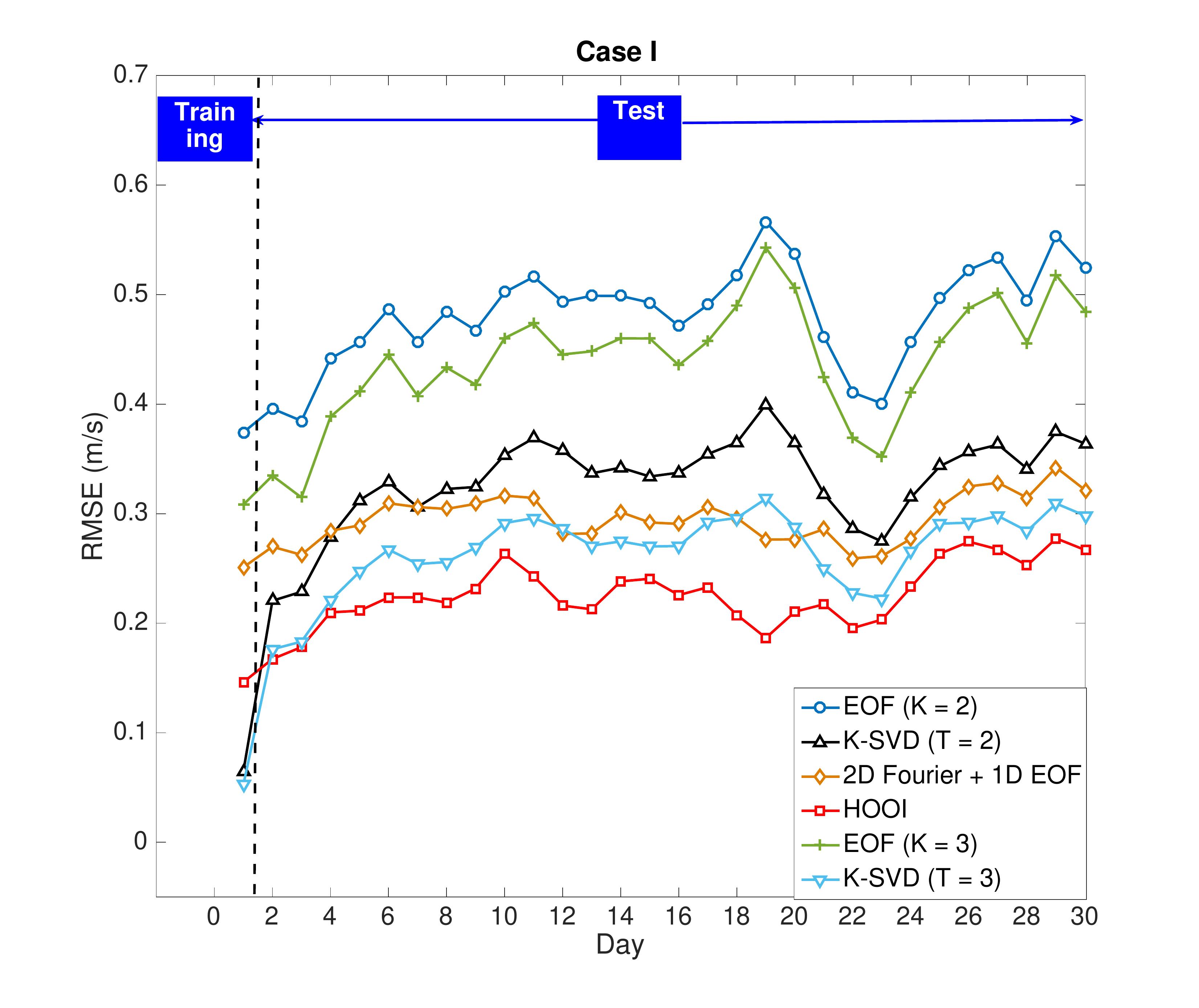}{\reprintcolumnwidth}{(a)} \label{fig:FIG5a}
\fig{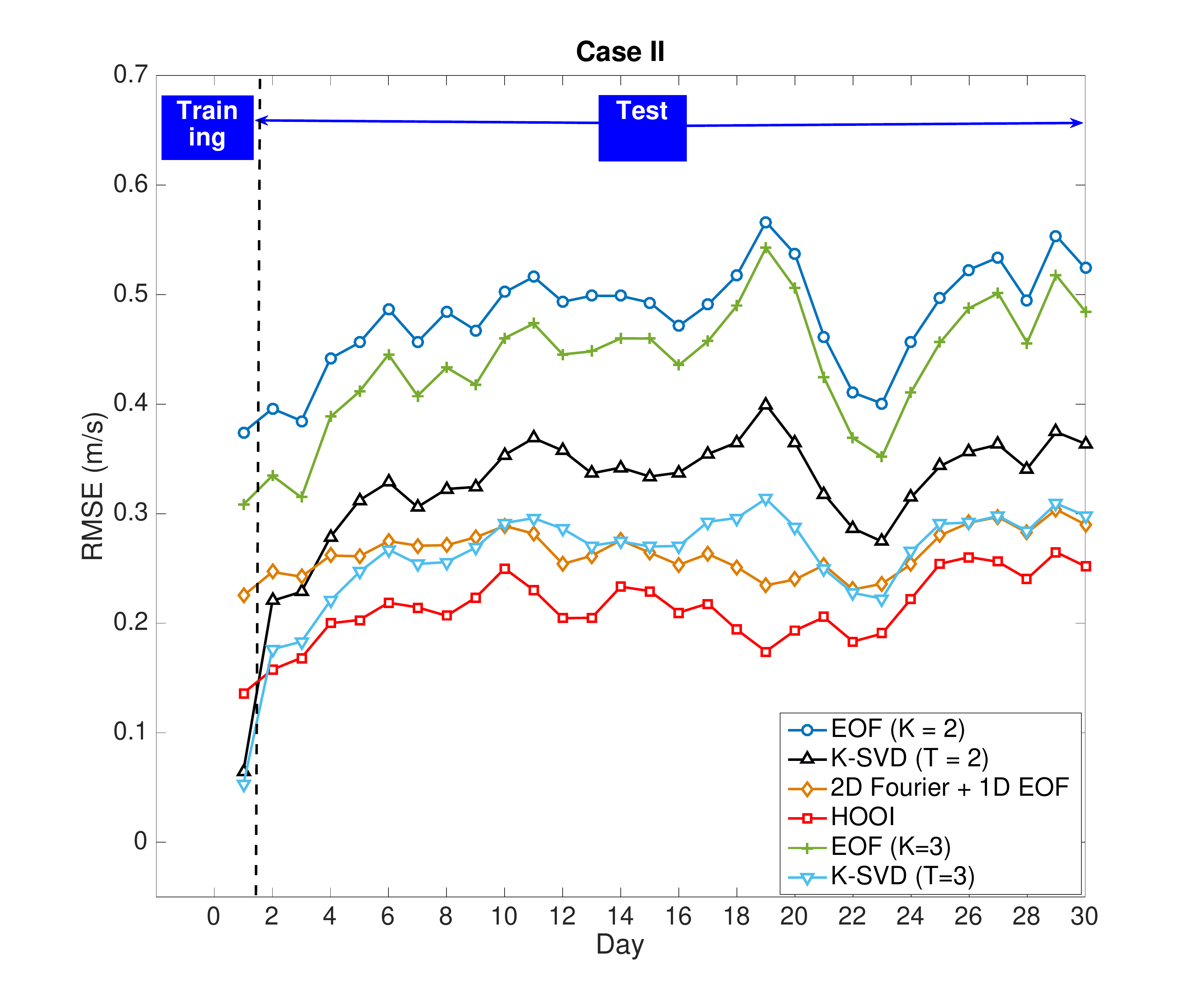}{\reprintcolumnwidth}{(b)}  \label{fig:FIG5b}
}
\caption{The RMSEs of different algorithms versus the training data and the test data under Case I (a) and Case II (b).}
\hrule
\end{figure*}

 \begin{figure*}[t]
\includegraphics[width= 6in]{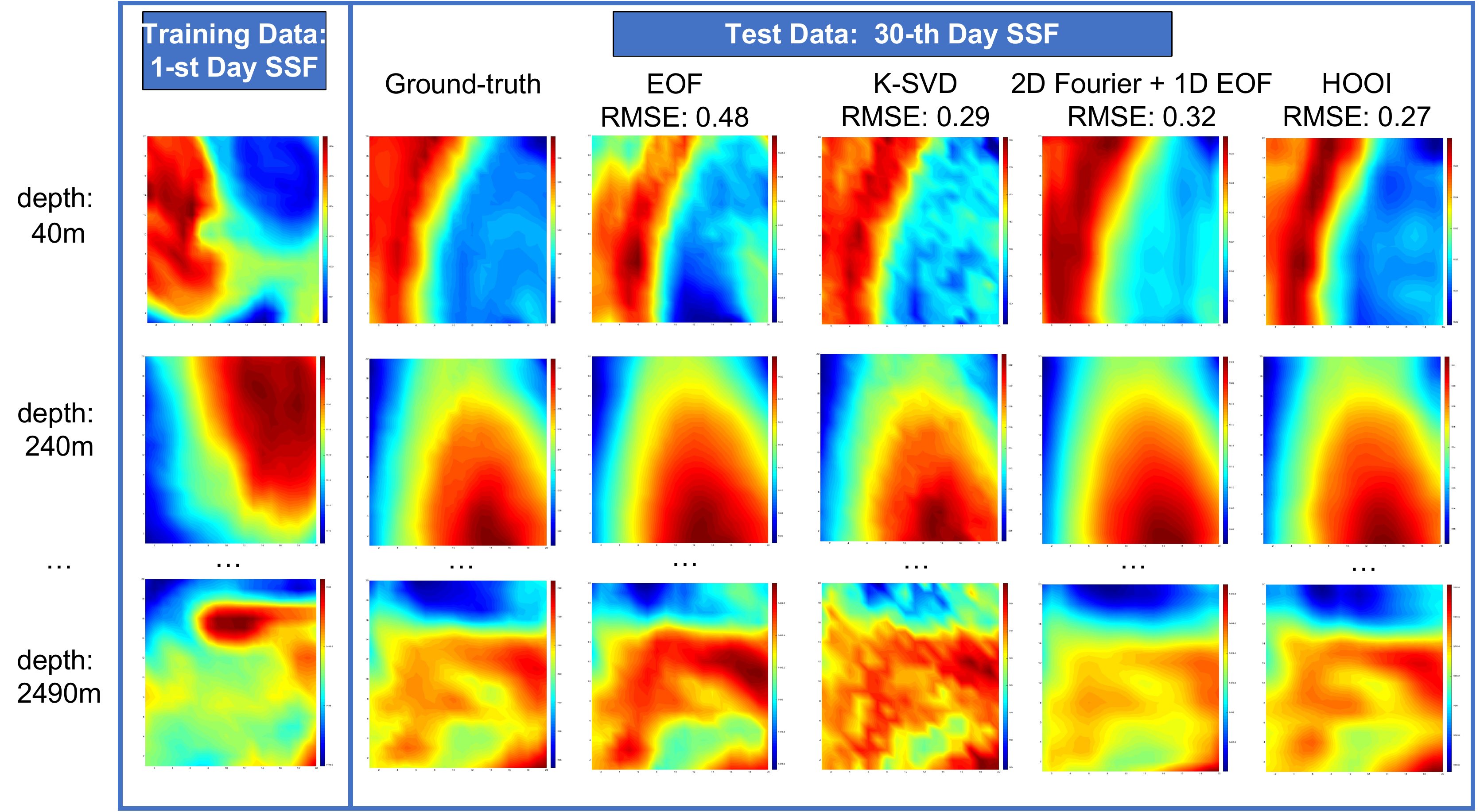}
\caption{Visual effects of the 3D SSF reconstruction for horizontal slices at depth $40$m, $240$m and $2490$m. The test data is the $30$-th day 3D SSF data, and the training data is the $1$-st day 3D SSF data. The hyper-parameters of the algorithms follow those in Case I, Table~\ref{tab:table2}. Particularly, the EOF-based method is with $K =3$, and the K-SVD-based method is with $T =3$. The tensor-based basis functions learnt from the HOOI algorithm give the best reconstruction performance.}
\label{fig:FIG6}
\raggedright
\hrule
\end{figure*}

\begin{figure}[!t]
\centering 
\includegraphics[width= 1\reprintcolumnwidth]{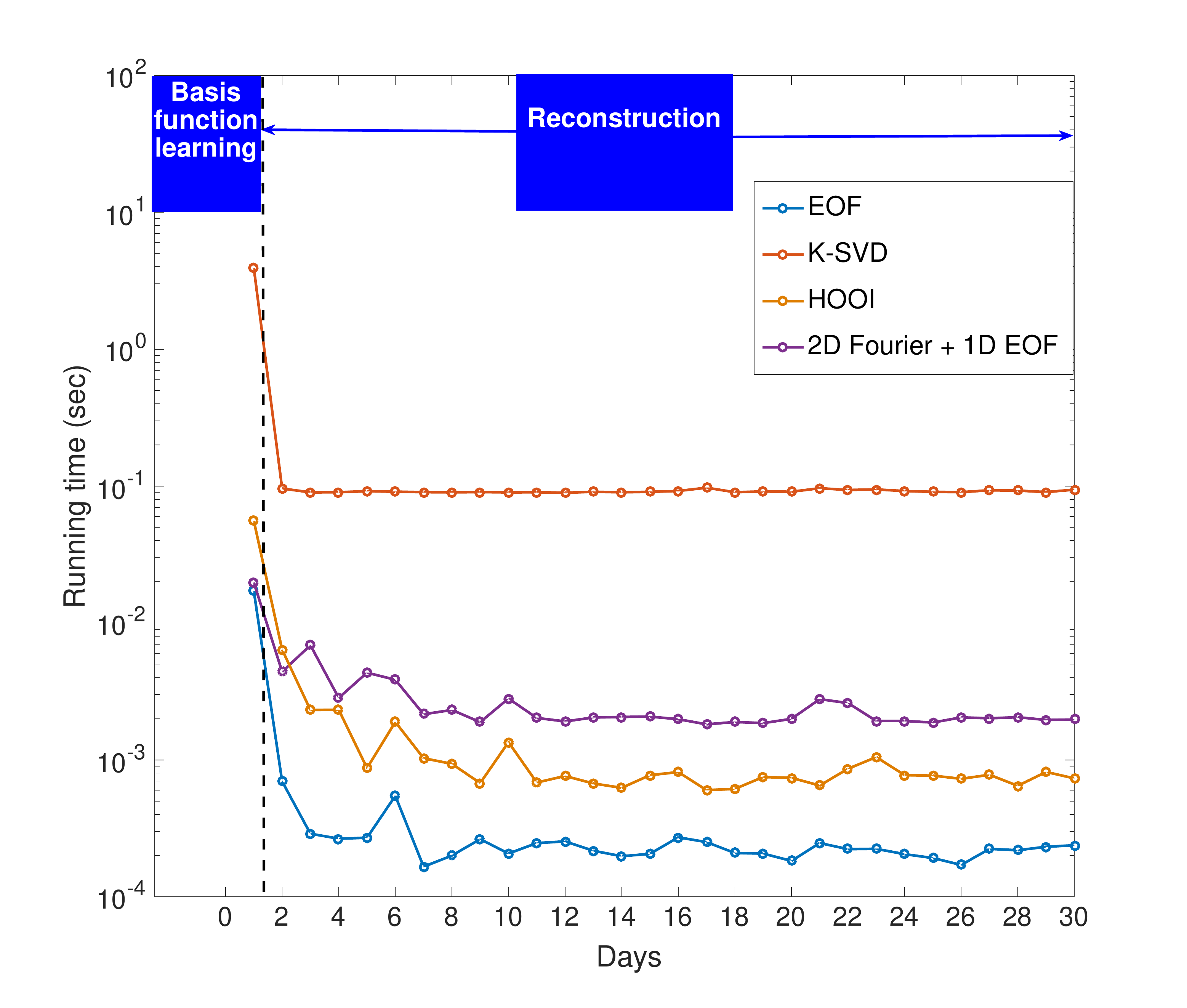}
\caption{The running time of different algorithms in basis function learning phase and SSF reconstruction phase.}
\label{fig:FIGN9}
\end{figure}

In Fig.~\ref{fig:FIG5a} and Fig.~\ref{fig:FIG5b}, we present the RMSEs versus the training data and the test data under the two cases (see Table \ref{tab:table2}), respectively.  {\color{black} Although} the tensor-based basis functions (learnt from HOOI algorithm) are with the minimal number of representation coefficients, the associated test RMSEs are always smaller than those of  benchmarking algorithms. On the other hand, in the training phase, the basis functions from K-SVD algorithm give the lowest RMSE, while the RMSE of tensor-based basis functions is the second lowest one. However, in the test phase, the RMSEs of the K-SVD-based basis functions quickly increase  {\color{black} and} are always larger than those of the HOOI-based basis functions.  This  {\color{black} observation indicates} that the K-SVD method overfits the training data, while the HOOI algorithm exhibits much better generalization performance when dealing with unseen data. Finally, although the classical basis functions (2D Fourier + 1D EOF) are with the same number of representation coefficients as the tensor-based counterpart, the resulting RMSEs are higher in both training phase and test phase. The EOF-based method shows the worst performance in 3D SSF representation. {\color{black} Note that a large reduction in the RMSEs of EOF-based method occurs on days 20-24, since  the sound speed variations (in the vertical domain) of the $23$-th day are more similar to those of the $1$-st day than other nearby days.} These results show the effectiveness of data-driven approach in representation learning, since the three factor matrices (that contain basis functions) are learnt from data when adopting HOOI algorithm (see Table~\ref{tab:prop}).

In Fig.~\ref{fig:FIG6}, we present the reconstructed SSF horizontal slices at different depths for the $30$-th day test data, with the $1$-st day training data serving as the reference. First, although the test data is very different from the training one, all the basis functions, which are learnt from the training data, can represent the test data to different extents of accuracy. Second,  {\color{black}  the tensor-based basis functions} learnt from the HOOI algorithm give the  best SSF reconstruction. Finally, the overfitting issue of K-SVD method can  be also observed.

\subsubsection{Running Time}

In Fig.~\ref{fig:FIGN9}, we present the running time of different algorithms  in basis function learning phase (training phase) and SSF reconstruction phase (test phase). The hyper-parameters of the algorithms follow those in Case I, Table II. Particularly, the EOF-based method is with $K = 3$, and the K-SVD-based method is with $T = 3$. {\color{black} Note that in the first day, the training algorithms of different methods are performed, which cost much more time than those of test process. Thus, the running time of test process (corresponding to the 2-30 days) is much less than that of the first day. This is the reason that a decrease of running time occurs.   For the 2-30 days (corresponding to test process), the fluctuation of running time  is mostly at the order of $10^{-3}$ sec, which can be viewed as the negligible systematic biases of computer hardware.}

 {\color{black}  The K-SVD-based method} costs the most time in two phases, since both the basis function learning and SSF reconstruction demand iterative algorithms. The classical basis functions (i.e., EOF and 2D Fourier + 1D EOF) are with closed-form expressions in two phases, thus costing much less time. On the other hand, the HOOI algorithm for basis function learning needs iterative updates, thus costing the second most time. However, the  {\color{black}  reconstructions} using the tensor-based basis functions has a closed-form expression  (see  {\color{black}  Eq.~\eqref{eq17}})  and thus is very fast. In the test phase, note that the running time using the 2D Fourier + 1D EOF basis function is slightly higher than that using the HOOI-based basis functions.  {\color{black}  The reason is that} the Fourier basis functions have introduced the computations of complex numbers for reconstruction, while the HOOI-based reconstruction only has computations of real numbers.

\subsubsection{The Number of Coefficients Required for Accurate SSF Reconstruction}

In this subsection, we evaluate the representation capabilities of different basis functions in terms of their hyper-parameters (or equivalently the number of representation coefficients). The mapping between the values of hyper-parameters and the number of representation coefficients {\color{black} is shown in Table~\ref{tab:table3} of Appendix~\ref{appendix-c}}.

\begin{figure*}[t]
\baselineskip=12pt
\figline{
\fig{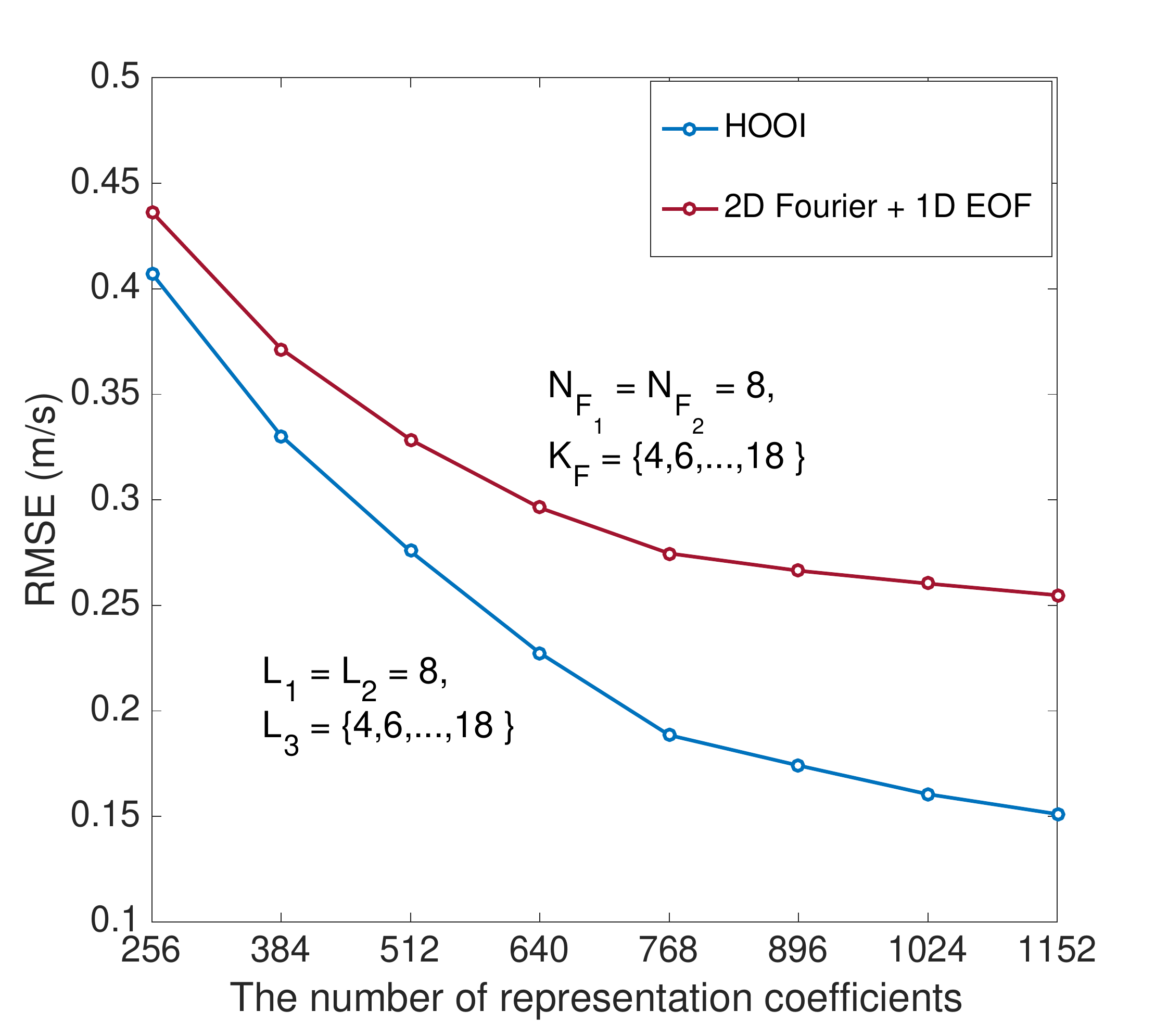}{\reprintcolumnwidth}{(a)} \label{fig:FIG6a}
\fig{Figure10a-eps-converted-to.pdf}{\reprintcolumnwidth}{(b)}  \label{fig:FIG6b}
}
\caption{\label{fig:FIGN10}The average test RMSEs versus different number of representation coefficients: (a) varying the value of $L_3$ or $K_F$; (b)  changing the values of $\{L_1, L_2 \}$ or $\{N_{F_1}, N_{F_2}\}$.}
\hrule
\end{figure*}

\begin{figure}[!t]
\includegraphics[width=\reprintcolumnwidth]{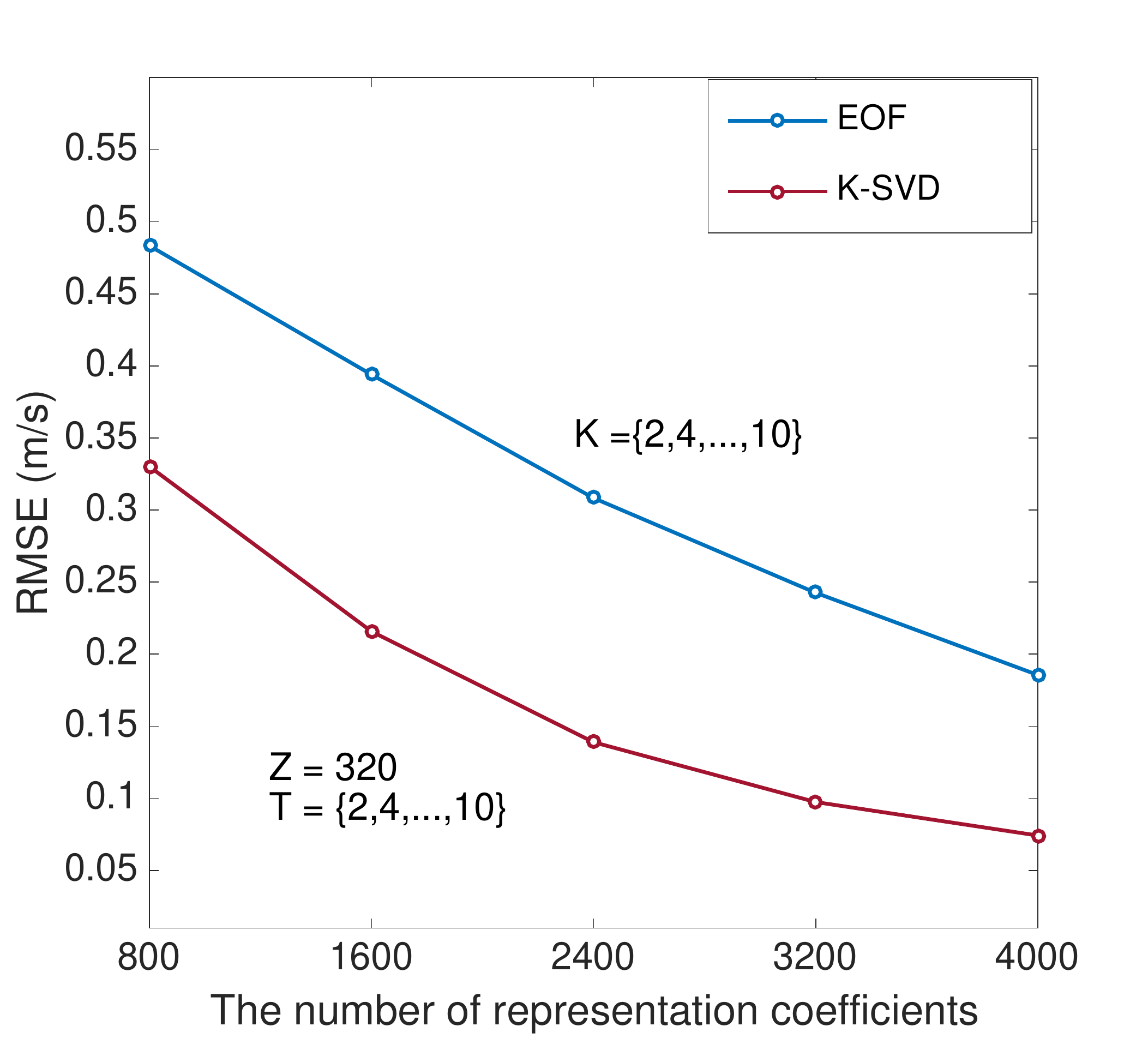}
\caption{\label{fig:FIG7}{ The average test RMSEs versus different number of representation coefficients for EOFs   and K-SVD-based basis functions. }}
\end{figure}

\begin{figure}[!t]
\includegraphics[width=\reprintcolumnwidth]{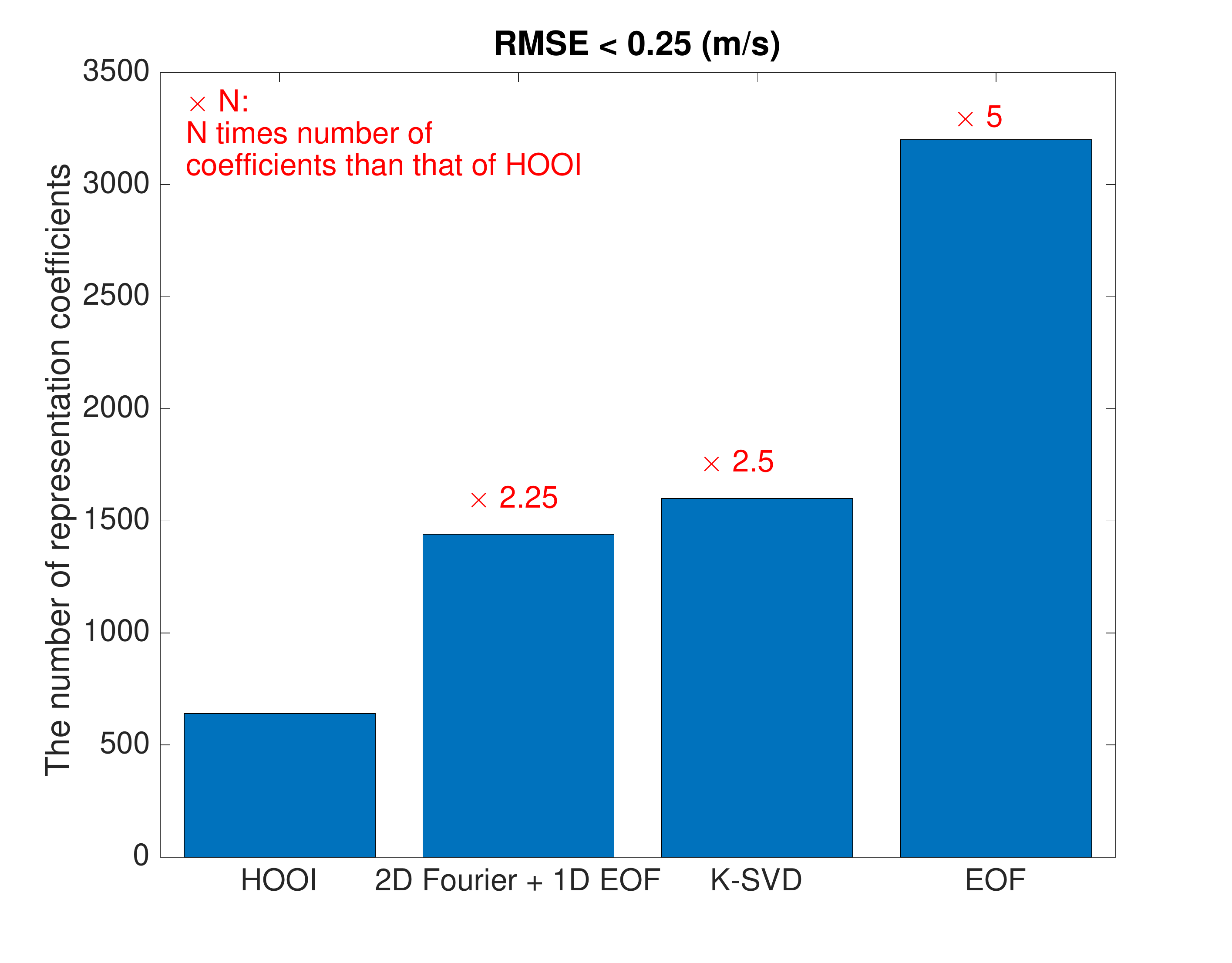}
\caption{\label{fig:FIG8}{The number of representation coefficients required for different basis functions that give the average test $\text{RMSE} < 0.25 $(m/s).}}
\end{figure}

{\color{black} In Fig.~\ref{fig:FIG6a}, by varying the value of $L_3$, which determines the vertical resolution of 3D SSF representation,  the average test RMSEs (over $29$ test days) of tensor-based basis functions (HOOI) are presented. Meanwhile, by changing the values of $K_F$, the average  test RMSEs (over $29$ test days) of the classical basis functions (2D Fourier + 1D EOF) are  provided.} On the other hand, {\color{black} changing the values of $\{L_1, L_2 \}$ or $\{N_{F_1}, N_{F_2}\}$  affects the horizontal resolution of 3D SSF representation. The average test RMSEs  of the two types of basis functions are shown  in Fig.~\ref{fig:FIG6b}.} The tensor-based basis functions give much better reconstruction accuracies than those of the classical basis functions, showing the superiority of tensor tools in 3D Ocean signal processing. In addition, {\color{black} the RMSE decreases more in total due to the horizontal resolution increase, but achieves overall lower value with high vertical resolution.}  For comparison, in Fig.~\ref{fig:FIG7}, we present the average test RMSEs of the EOF-based and the K-SVD-based basis functions by setting their hyper-parameters to different values. Note that in Fig.~\ref{fig:FIGN10} and Fig.~\ref{fig:FIG7}, we show the RMSEs versus the number of representation coefficients (see the mapping in Table~\ref{tab:table3} of  Appendix~\ref{appendix-c}), to show the effectiveness of different basis functions more straightforwardly. 

In Fig.~\ref{fig:FIG8}, we show the number of representation coefficients required for different basis functions such that the average test $\text{RMSE} < 0.25$(m/s) can be achieved.  {\color{black}  The tensor-based method requires} the minimal number of representation coefficients, showing its high expressive power in representing 3D SSF. Other basis functions  require at least  {\color{black}  two times} number of coefficients than the tensor-based counterpart. 

{\color{black}
\subsection{Learning Basis Functions From Multiple 3D SSFs Across Different Seasons}
In this subsection,  we assess the performance of tensor-based basis functions learnt from multiple 3D SSFs across different seasons in one year.

\textbf{3D SSF Data:}  The 3D South China Sea (SCS)  SSFs $\{\bc X_t  \in \mathbb R^{13 \times 13 \times 37 }\}$ in the year 2020 are analyzed in this subsection. The data was provided by National Marine Data Center  (\url{http://mds.nmdis.org.cn/}). The considered spatial area  ($152 \text{km} \times 152 \text{km} \times 2 \text{km}$) is with the same longitudes and latitudes as those in Section~\ref{sec:v-a}, while with lower horizontal and vertical resolutions. 

\textbf{Training Data and Test Data:} The 3D SSFs used for training are from four months in the year 2020, namely, February, May, August, and November.  Note that these four months correspond to four seasons in one year. In each month, the 3D SSFs of the first three days are selected for the training purpose.  Consequently, twelve 3D SSFs across different seasons give the training data. To evaluate the representation performance across four seasons, one-week 3D SSFs in each of these four months are employed as the test data. As a result, there are twenty-eight unseen 3D SSFs used for testing.  Note that the selected test 3D SSFs do not contain the training 3D SSFs.

\textbf{Baselines and Performance Metric:}  Following those in Section~\ref{sec:v-a}.

\subsubsection{Reconstruction Error under The Similar Number of Representation Coefficients}

We compare the RMSEs of different algorithms given the similar number of representation coefficients in Fig.~\ref{fig:FIGM1}. The hyper-parameters of these algorithms were set to let their corresponding representation coefficients have similar numbers, as shown in Table~\ref{tab:tablenew}. From Fig.~\ref{fig:FIGM1}, the RMSEs of  tensor-based basis functions (labeled as M-HOOI), whose coefficient number is the smallest,  are lower than other benchmarking algorithms across different seasons in most cases.  These results show that the tensor-based basis functions jointly learnt from multiple 3D SSFs are capable of accurate yet reduced-order representation for the 3D SSFs in a long span of time.

 \begin{table}[!t]

 \caption{ \label{tab:tablenew}The hyper-parameters and the number of representation coefficients for different algorithms.}
 \begin{ruledtabular}
\begin{tabular}{|c|c|c|c|c|}
Algorithms &
  M-HOOI &
  EOF &
  K-SVD &
  \begin{tabular}[c]{@{}c@{}}2D Fourier \\ + 1D EOF\end{tabular} \\ \hline
Hyper-parameters &
  \begin{tabular}[c]{@{}c@{}}$L_1 = 6,$\\ $L_2 = 6,$\\ $L_3 = 8$\end{tabular} &
  $K = 2$ &
  \begin{tabular}[c]{@{}c@{}}$T = 2,$\\ $Z = 40$\end{tabular} &
  \begin{tabular}[c]{@{}c@{}}$N_{F_1 }=6, $\\ $N_{F_2 }=6,$\\ $K_F = 8$\end{tabular} \\ \hline
\begin{tabular}[c]{@{}c@{}}The Number of \\ Representation\\ Coefficients\end{tabular} &
  $288$ &
  $338$ &
  $338$ &
  $288$ \\
\end{tabular}
\end{ruledtabular}
\end{table}

\begin{figure}[!t]
\centering 
\includegraphics[width= \reprintcolumnwidth]{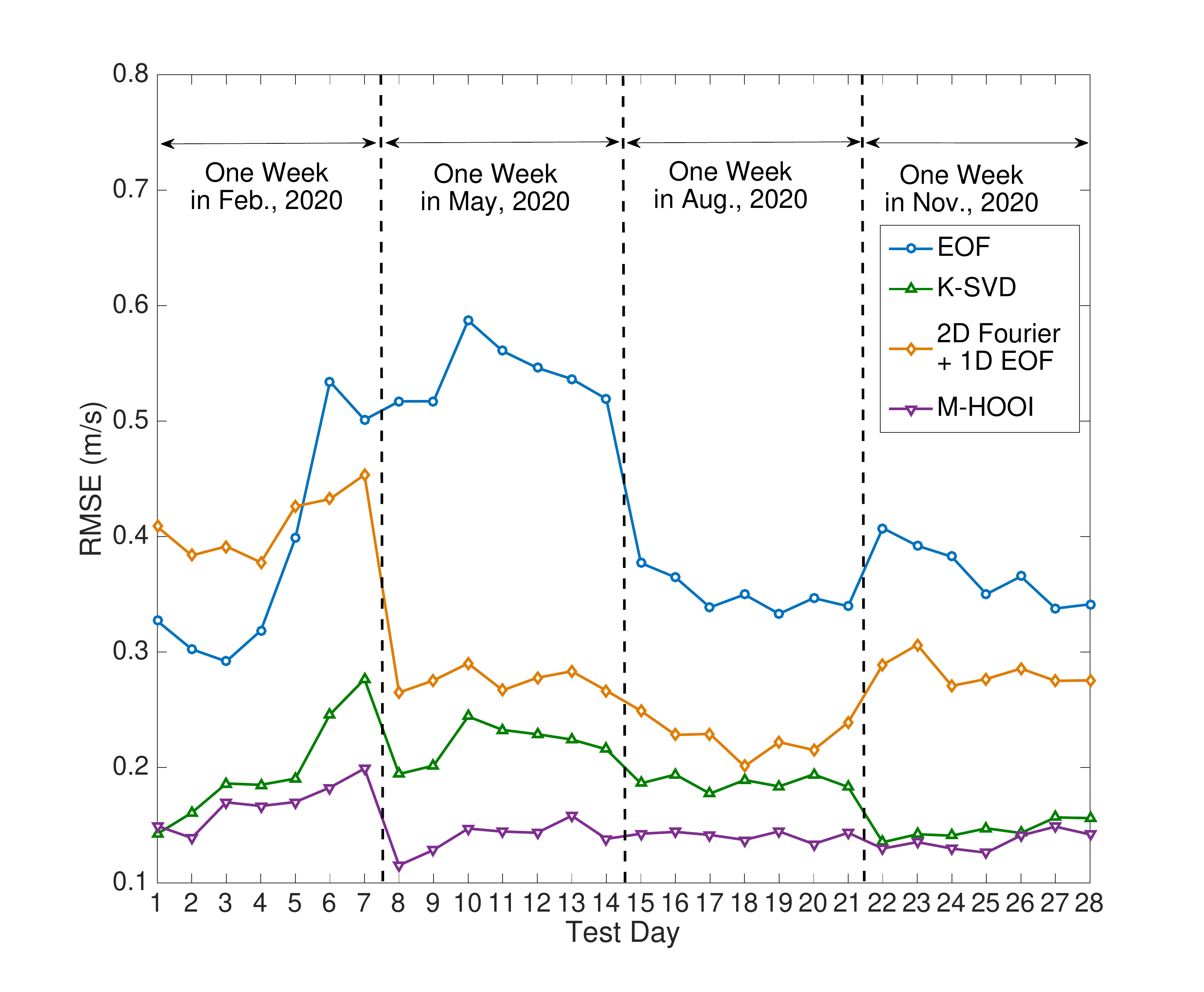}
\caption{The RMSEs of different algorithms versus the test data across different months.}
\label{fig:FIGM1}
\end{figure}

\begin{figure*}[t]
\baselineskip=12pt
\figline{
\fig{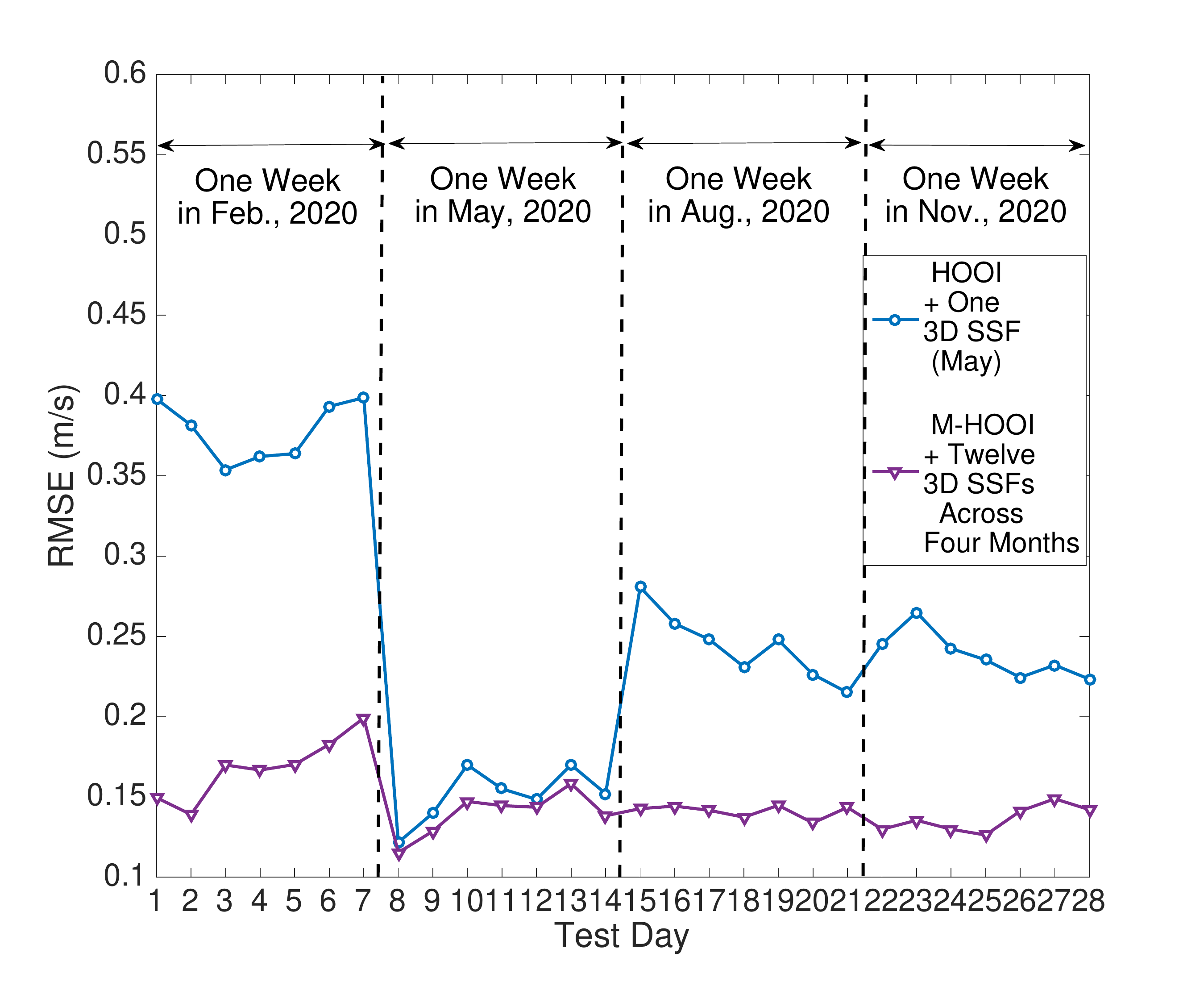}{\reprintcolumnwidth}{(a)} \label{fig:FIGM2a}
\fig{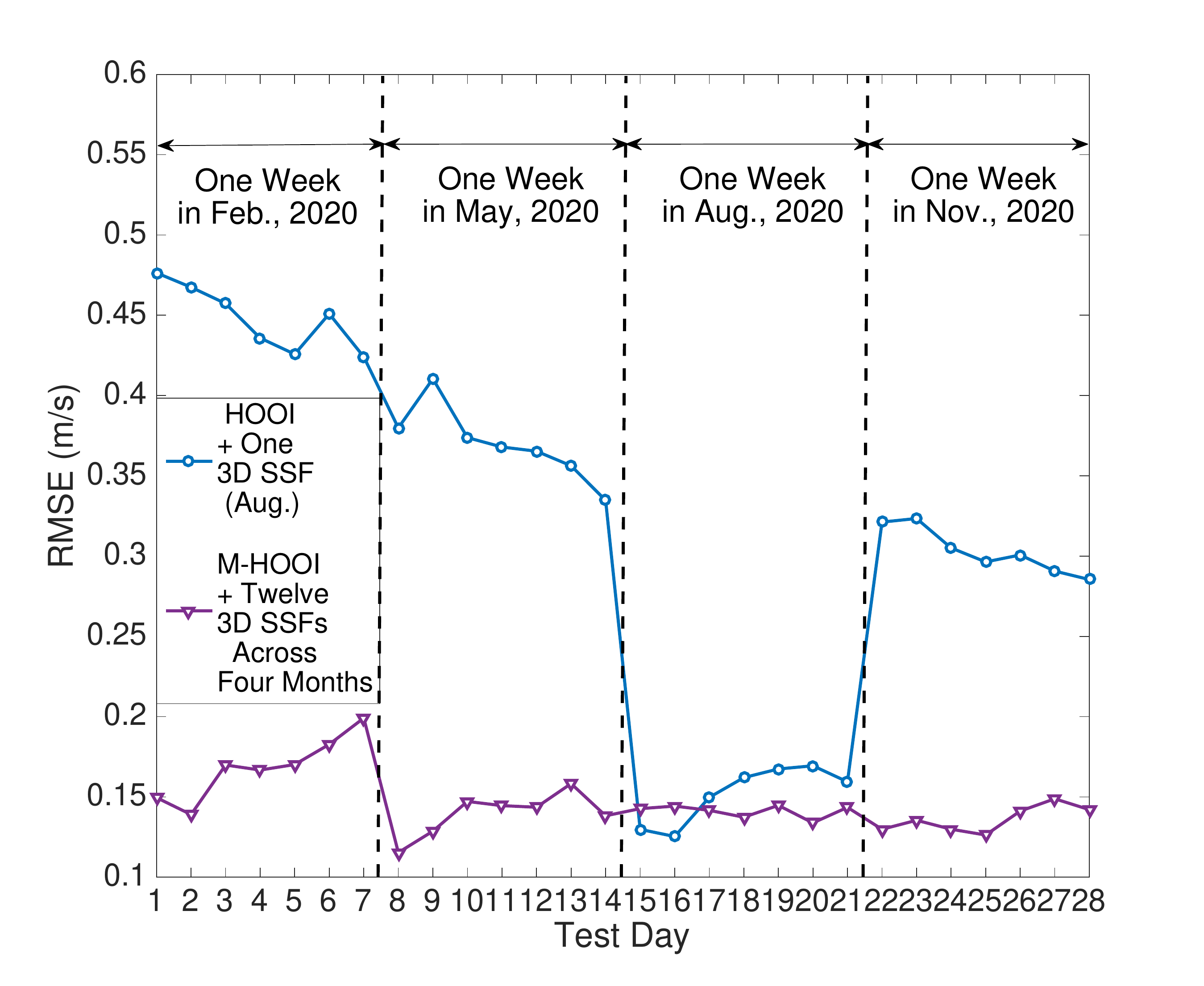}{\reprintcolumnwidth}{(b)}  \label{fig:FIGM2b}
}
\caption{\label{fig:FIGM2} The RMSEs of tensor-based basis function learning schemes using one 3D SSF and multiple 3D SSFs.}
\hrule
\end{figure*}

\subsubsection{Reconstruction Error using One 3D SSF or Multiple 3D SSFs}
In Fig.~\ref{fig:FIGM2}, we compare the performance of the basis functions learnt from one 3D SSF (labeled as HOOI + one 3D SSF (Month)) and those learnt from multiple 3D SSFs (labeled as M-HOOI + Twelve 3D SSFs Across Four Months). Note that the basis functions compared in this subsection are with the same order, i.e., $L_1 = 6, L_2 =6 , L_3 = 8$. Particularly, in Fig.~\ref{fig:FIGM2a} and Fig.~\ref{fig:FIGM2b}, the $1$-st day 3D SSF of May, 2020 and  the $1$-st day 3D SSF of Aug., 2020 are used as the training data for HOOI algorithm respectively.  In contrast, the M-HOOI algorithm learns the basis functions from twelve 3D SSFs across four months in 2020 (as introduced at the beginning of this subsection).

Fig.~\ref{fig:FIGM2} shows that the basis functions learnt from one 3D SSF in a particular month can well represent the unseen 3D SSFs in one week of that month. However, for the 3D SSFs in other months, their performance degrades. On the other hand, the basis functions jointly learnt from twelve 3D SSFs across four seasons give much  lower RMSEs in most test cases. These results show that using more training 3D SSFs across different seasons,  the learnt tensor-based  basis functions can realize long-term effective 3D SSF representation.  Note that this advantage is  at the cost of more historical training data from different months/seasons in the observed sea area, which might not be available in some applications.}

\section{Conclusions and Future Directions}
\label{sec:6}

In this paper, by treating the 3D SSF data as a third-order tensor, a tensor-based basis function learning framework was introduced. Under this framework,  the classical basis functions using EOFs and Fourier basis functions can be treated as the special cases. Relying on the Tucker tensor decomposition format, the HOOI algorithm  and M-HOOI algorithm were introduced to learn the effective basis functions from one 3D SSF and multiple 3D SSFs in a data-driven fashion. Numerical results using SCS 3D SSF data have showcased the excellent performance of tensor-based basis functions in terms of both reconstruction accuracy and running time. 

The HOOI and M-HOOI algorithms exemplify  the use of tensor tools (e.g., the proposed tensor-based basis function learning framework) in multi-dimensional ocean signal processing. In future research, it is possible to obtain better basis functions by imposing informative equality/inequality constraints that incorporate more information of 3D SSF, e.g., the fact that sound speeds vary much more significantly in the shallow ocean than those in the deep ocean.   {\color{black}  The integration} of {\it physical science} and {\it data science} will bring us closer towards the {\it Universal Representation (UR)} of ocean signals.

\section{Acknowledgement}
The authors would like to thank the Institute of Oceanology, Chinese Academy of Sciences, and National Marine Data Center for providing the Ocean 3D SSF data for analysis. This work was supported  in part by the National Natural Science Foundation of China under Grant 62001309 and Grant 62071429,  and in part by Shanghai Aerospace Science and Technology Innovation Foundation (Grant No. SAST2020-034).

\appendix
\section{The Proof of Proposition 1}
\label{appendix-a}

Here we show that the solution of problem \eqref{eq221}  gives the classical  EOFs for 2D SSF (see Section~\ref{Sec II-A}). 

Using variable substitution, problem \eqref{eq221} can be equivalently expressed as  {\color{black}  follows}
\begin{align}
    &\min_{\bc S, \mb B^{(3)}} \left \Vert \bc X -  \bc S \times_1 \mb I_{M} \times_2 \mb I_{N}  \times_3   \mb B^{(3)} \right \Vert_\F^2, \nonumber\\
    & \text{s.t.} ~~ \bc S \in  \mathbb R^{M \times N  \times K}, \nonumber \\
       & ~~~~~ \mb B^{(3)} \in \mathbb R^{I \times K}, ~[\mb B^{(3)}]^\T \mb B^{(3)} = \mb I_{K}.
    \label{eqa1}
\end{align}
After substituting   {\color{black}  Eq.~\eqref{eq17} into Eq.~\eqref{eqa1}}, the remaining problem for solving $\mb B^{(3)}$ becomes
\begin{align}
    & \max_{\mb B^{(3)} \in \mathbb R^{I \times K}} \left  \| \left[\mb B^{(3)}\right]^\T \mb C_{(3)} \right \|_\F^2, \nonumber \\
       & \text{s.t.}  ~~ [\mb B^{(3)}]^\T \mb B^{(3)} = \mb I_{K}, 
           \label{eqa2}
\end{align}
where 
\begin{align}
    \mb C_{(3)} = \mb X_{(3)} ( \mb F_2 \otimes \mb F_1).
\end{align}
Note that the mode-$3$ unfolding matrix $\mb X_{(3)}$ is equal to the matrix $\mb X^\text{u}$ defined in Section~\ref{sec:ii-c} (see Remark 3). Therefore, the solution of problem \eqref{eqa2} is
\begin{align}
\mb B^{(3)} = \left [ \mb e_1, \mb e_2, \cdots, \mb e_{K_F} \right],
\label{eqa3}
\end{align}
where vector $\mb e_k$ is the $k$-th leading left singular vectors of $\mb X^\text{u}$. According to the definition of EOFs, it can be concluded that 
\begin{align}
\mb B^{(3)}  = \mb E_{K}.
\label{eqa4}
\end{align}
Then, the proof of Proposition 1  {\color{black}  is completed. }

\section{The Proof of Proposition 2}
\label{appendix-b}

Similarly, using variable substitution, problem \eqref{eq22} can be equivalently expressed as follows:
\begin{align}
    &\min_{\bc S, \mb B^{(3)}} \left \Vert \bc X -  \bc S \times_1 \mb F_1 \times_2 \mb F_2  \times_3   \mb B^{(3)} \right \Vert_\F^2, \nonumber\\
    & \text{s.t.} ~~ \bc S \in  \mathbb R^{N_{F_1} \times N_{F_2} \times K_F}, \nonumber \\
       & ~~~~~ \mb B^{(3)} \in \mathbb R^{I \times K_F}, ~[\mb B^{(3)}]^\T \mb B^{(3)} = \mb I_{K_F}.
    \label{eqb1}
\end{align}
After substituting   {\color{black}  Eq.~\eqref{eq17} into Eq.~\eqref{eqb1}}, the remaining problem for solving $\mb B^{(3)}$ becomes
\begin{align}
    & \max_{\mb B^{(3)} \in \mathbb R^{I \times K_F}} \left  \| \left[\mb B^{(3)}\right]^\T \mb C_{(3)} \right \|_\F^2, \nonumber \\
       & \text{s.t.}  ~~ [\mb B^{(3)}]^\T \mb B^{(3)} = \mb I_{K_F}, 
           \label{eqb2}
\end{align}
where 
\begin{align}
    \mb C_{(3)} = \mb X_{(3)} ( \mb F_2 \otimes \mb F_1).
\end{align}
From  {\color{black} Eq.~\eqref{eqa3} and  Eq.~\eqref{eqa4}}  in Appendix~\ref{appendix-a},  we can conclude that the solution of problem \eqref{eqb2} is
\begin{align}
\mb B^{(3)}  = \mb E_{K_F}.
\end{align}
With $\mb B^{(1)}= \mb F_1, \mb B^{(2)} = \mb F_2, \mb B^{(3)}  = \mb E_{K_F}$, the proof of Proposition 2  {\color{black} is completed}.

\section{Representation Coefficients}
\label{appendix-c}
The mapping between the hyper-parameters and the number of representation coefficients for different algorithms is shown in Table~\ref{tab:table3}.

 \begin{table}[!t]

 \caption{ \label{tab:table3}The mapping between the hyper-parameters and the number of representation coefficients for different algorithms.}
 \begin{ruledtabular}
\begin{tabular}{|c|c|c|c|c|}
Algorithms &
  HOOI &
  EOF &
  K-SVD &
  \begin{tabular}[c]{@{}c@{}}2D Fourier \\ + 1D EOF\end{tabular} \\ \hline
Hyper-parameters &
  \begin{tabular}[c]{@{}c@{}}$L_1,$\\ $L_2,$\\ $L_3$\end{tabular} &
  $K$ &
  \begin{tabular}[c]{@{}c@{}}$T,$\\ $Z$\end{tabular} &
  \begin{tabular}[c]{@{}c@{}}$N_{F_1}, $\\ $N_{F_2},$\\ $K_F$\end{tabular} \\ \hline
\begin{tabular}[c]{@{}c@{}}The Number of \\ Representation\\ Coefficients\end{tabular} &
  $L_1L_2L_3$ &
  $KMN$ &
  $TMN$ &
  $N_{F_1}N_{F_2} K_F$ \\
\end{tabular}
\end{ruledtabular}
\end{table}

\section{The Requirement of 3D SSFs}
\label{appendix-d}

{\bf The size of the 3D sea area.}   If the considered sea area is very small (e.g., 100m $\times$ 100m $\times$ 50m), it is very likely that the corresponding 3D SSF tensor has many front slices being the same. In other words, the 3D SSF tensor can be simply   treated as  stacking the same  SSP matrix multiple times. In this case, tensor decomposition reduces to matrix decomposition, and thus the proposed tensor-based method is with nearly the same performance as those of matrix-based methods (e.g., EOF). Therefore, the proposed method has advantages when the considered 3D sea area is with horizontally spatial scale of $O(100\text{km})$ and vertically spatial scale of $O(1000\text{m})$ (e.g.,  160km $\times$ 160km $\times$ 3km). In such areas, mesoscale ocean dynamics (e.g., mesoscale eddy as shown in Fig. 5) will endue the 3D SSF a low-rank tensor structure, and thus ensure the superiority of the proposed method over classical matrix-based methods. 
 
 {\bf The topography of the seafloor.}  The sea floor is supposed to be even/flat such that the considered 3D SSF has values in each entry. Otherwise, the 3D SSF cannot be represented by a regular 3D tensor, for which advances in tensor completion\cite{panagakis2021tensor} might be explored as an interesting future research work. 
  
 {\bf The selection of sampling points.} The dense sampling interval  is useful for learning basis functions with better performance, since it provides more informative training data.  Usually, uniform sampling of 3D sea area is preferred since the resulting sampling points can preserve spatial-temporal correlations as much as possible.  Nevertheless, non-uniform sampling along the vertical dimension is viable (and also widely adopted in practice), since the sound speeds vary more significantly in the shallow ocean than those in the deep ocean. Thus, one can densely sample the shallow ocean while sparsely sample the deep ocean.

\bibliography{sampbib}

\end{document}